\documentclass[apj,iop]{emulateapj}

\usepackage{amsmath}
\usepackage{tabularx}
\usepackage{natbib}
\usepackage{float}
\usepackage[colorlinks=true,linkcolor=blue,citecolor=blue]{hyperref}%
\usepackage[all]{hypcap}
\usepackage{lineno}
\usepackage{footnote}

\DeclareUnicodeCharacter{2212}{-}

\newcommand{\teff}{$T_{\rm eff}$}
\newcommand{\tint}{$T_{\rm int}$} 

\newcommand{\co}{CO}
\newcommand{\meth}{CH$_4$}
\newcommand{\amon}{NH$_3$}
\newcommand{\cotwo}{CO$_2$} 
\newcommand{\water}{H$_2$O}

\newcommand{\tchem}{$t_{\rm chem}$}
\newcommand{\tmix}{$t_{\rm mix}$}
\newcommand{\kzz}{$K_{zz}$}

\newcommand{\orcid}[1]{\href{https://orcid.org/#1}{\includegraphics[width=10pt]{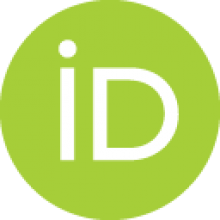}}}

\shorttitle{\texttt{PICASO 3.0}: 1D Climates}
\shortauthors{Mukherjee et al.}
\graphicspath{{./}{figures/}}
\begin{document}

\title{\texttt{PICASO 3.0}: A One-Dimensional Climate Model for Giant Planets and Brown Dwarfs }

%\correspondingauthor{Sagnick Mukherjee}
\email{samukher@ucsc.edu}

\author{Sagnick Mukherjee$^{1}$ \orcid{0000-0003-1622-1302},  Natasha E. Batalha$^{2}$ \orcid{0000-0003-1240-6844}, Jonathan J. Fortney$^{1}$ \orcid{0000-0002-9843-4354}, Mark S. Marley$^{3}$ \orcid{0000-0002-5251-2943}}
\affiliation{{$^1$}Department of Astronomy and Astrophysics, University of California, Santa Cruz, CA 95064, USA \\ 
{$^2$} NASA Ames Research Center, MS 245-3, Moffett Field, CA 94035, USA \\
{$^3$} Lunar and Planetary Laboratory, The University of Arizona, Tucson, AZ 85721, USA\\}

\begin{abstract}

Upcoming James Webb Space Telescope (JWST) observations will allow us to study exoplanet and brown dwarf atmospheres in great detail. The physical interpretation of these upcoming high signal--to--noise observations requires precise atmospheric models of exoplanets and brown dwarfs. While several one-dimensional and three-dimensional atmospheric models have been developed in the past three decades, these models have often relied on simplified assumptions like chemical equilibrium and are also often not open-source, which limits their usage and development by the wider community. We present a python-based one-dimensional atmospheric radiative-convective equilibrium model. This model has heritage from the Fortran-based code \citep{Marley96} which has been widely used to model the atmospheres of Solar System objects, brown dwarfs and exoplanets. In short, the basic capability of the original model is to compute the atmospheric state of the object under radiative--convective equilibrium given its effective or internal temperature, gravity and host--star properties (if relevant). In the new model, which has been included within the well-utilized code-base \texttt{PICASO}, we have added these original features as well as the new capability of self-consistently treating disequilibrium chemistry. This code is widely applicable to  Hydrogen-dominated atmospheres (e.g. brown dwarfs and giant planets).

\end{abstract}

\keywords{ Brown Dwarfs, exoplanets}
%%%%%%%%%%%%%%%%%%%%%%%%%%%%%%%%%%%%%%%%%%%%%%%%%%%%%%%%%%%%%%%%
\section{Introduction}\label{sec:intro}

There are three broad categories of substellar atmosphere modeling frameworks: one-dimensional (1D) physically self-consistent models, three-dimensional (3D) General Circulation Models (GCMs) and atmospheric retrieval. Atmospheric models enable the understanding of physical and chemical processes driving the behaviour of substellar atmospheres. When combined with observational data, they can be used to infer the physical state  of atmospheres. Additionally, the adiabat calculated by the atmospheric models  provides the upper boundary conditions of the interior evolutionary models of substellar objects like giant planets and brown dwarfs \citep{hubbard77,burrows97,chabrier00,saumonmarley08}. Although substellar atmospheres generally constitute a  small fraction ($\sim$ 1\%) of the total mass of their bodies, their radiative properties control the overall cooling of objects throughout their  evolution. Therefore, atmospheric models are crucial for understanding both physical (e.g. climate and chemistry) and global properties of substellar objects (e.g. radius/luminosity evolution over time).

A variety of 1D radiative--convective--thermochemical equilibrium models of exoplanetary and brown dwarf atmospheres have been developed \citep[e.g.][]{Marley96,barman01,fortney2005comp,fortney08,marley21,Philips20,Piskorz_2018,goyal2020}. These models originate both from stellar atmosphere modeling codes \citep[e.g.][]{barman01,sudarsky03,seager1998} and planetary atmospheric modeling codes \citep[e.g.][]{Marley96,baudino15}. These models rely on iterating to a self-consistent radiative-convective-thermochemical equilibrium (RCTE) solution. Self-consistency here means that the model iterates on all components of the atmospheric structure simultaneously such that they are physically consistent with each other (e.g., chemical abundances are consistent with the current temperature profile). The models do not capture 3D dynamics, but instead are able to incorporate the effects of radiative/convective energy transport, chemistry, and clouds.
These 1D models are also computationally efficient, which enhances their exploratory power significantly over 3D circulation models. For example, \citep[e.g.][]{zhang21} used the \texttt{SONORA BOBCAT} grid of models, generated using the 1D RCTE model described in \citet{marley21}, to infer the physical properties of 55 T-dwarfs in an uniform analysis. The computational speed of these 1D self-consistent models along with their ability to treat atmospheric physics and chemistry self-consistently makes them a  powerful tool for interpreting atmospheres of planets and brown dwarfs.

Besides the model used in \citet{marley21}, several other independent 1D RCTE models have been widely used in the literature \citep[e.g.][]{tremblin15,Philips20,gandhi17,burrows08,marley21,malik17}. For example, the \texttt{GENESIS} code from \citet{gandhi17} was used for the high significance detection of various Carbon and Nitrogen bearing species in the atmosphere of HD 209458b by \citet{giacobbe21}. The \texttt{ATMO} model \citep{tremblin15,Philips20} was used by \citet{goyal18} to perform a uniform fitting analysis of the transmission spectra of 10 hot Jupiters which led to important conclusions about their atmospheric composition.  Despite the many insights gained from these models, their simplicity often makes them insufficient to explain a variety of atmospheric spectra \citep[e.g.][]{noll97,oppenheimer98,saumon03,golimowski04,geballe09,sorahana2012,leggett12,Miles20}. For example, \citet{Miles20} found the presence of disequilibrium chemistry in a series of late T and early Y-dwarfs by measuring their significantly enhanced photospheric CO abundances. In another example, \citet{zhang21}  concluded that presence of clouds can lead to much better overall fit to the available infrared spectra of 55 brown dwarfs with a uniform analysis using the \texttt{SONORA} cloudless grid of models. This motivates adding physical complexity to 1D RCTE models in order to enable better interpretation of upcoming data, for example, from \textit{JWST}.

Models for exoplanetary and brown dwarf atmospheres have been developed in three-dimensions (3D) \citep[e.g.][]{showman08,showman20,tan21,menou09,roman19,Wolf_2022,Lee_2021} as well. The 3D models work to capture the detailed global dynamics like horizontal transport/winds in the  atmospheres along with other atmospheric components like radiation and clouds. However, these models are computationally intensive which limits their power of exploration of the vast parameter space in question. Moreover, in order to make the models computationally feasible,  they rely on approximations such as gray atmospheres in the radiative transfer calculations \citep[e.g.][]{tan21,tanandshowman19,menou09,tan22}. As a result, 3D models are unable to resolve some atmospheric properties predicted by 1D models -- like detached convective zones in brown dwarfs and giant planets.

Atmospheric retrievals are also widely used to interpret observational data. These studies aim to recover the atmospheric state of any planet/brown dwarf based on the observed spectral data in a Bayesian framework \citep[e.g.][]{madhu18,line17,burningham17}. The free parameters of interest typically include the atmospheric temperature-pressure ($T(P)$) profile, chemical abundances and cloud properties \citep{line17,burningham17,burningham21,mukherjee20,taylor21}. The multi-dimensional parameter space of these free parameters is sampled by repeatedly calculating model observables (e.g., spectra) with simple and computationally fast forward models. Comparison of these modeled observables with the observed data helps in retrieving likelihood distributions for all the retrieved free parameters.  However, these retrieved properties are often not required to have physical constraints. For example, the retrieval analysis of hot Jupiter, WASP-18 b, was predicted to have a thermal inversion and relatively high metallicity (283$\times$Solar), corresponding to a CO abundance of $>$10\% \citep{sheppard17}. Later using a self-consistent grid of models, \citet{arcangeli18} showed that this planet is more likely to have a solar metallicity and hot dayside, which is more consistent with what is expected from a hot Jupiter with little heat redistribution.  This means that even though retrieval studies are effective in inferring the various atmospheric properties like abundances and temperature structures, they are not equipped to provide the physical interpretations of why the atmosphere is in the retrieved state, and they often result in unphysical solutions. Therefore, physically motivated self-consistent models are required to understand the physical and chemical processes that drive the structure of planetary and substellar atmospheres.

In this work, we focus on 1D physically motivated radiative--convective--thermochemical equilibrium models for substellar atmospheres. In addition to include complexities like disequilibrium chemistry within these models, another important modification we must make to these codes, is to transform them to open sourced, ``FAIR'' codes (findable, accessible, interpretable, and reproducible). Doing so will enable the community to interpret the upcoming influx of data from missions like JWST \citep{JWSTERO}. Currently, there are a handful of 1D climate models that are open source (e.g., \texttt{HELIOS} \citep{malik17,malik19}, and \texttt{TLUSTY} \citep{hubeny88,hubeny03,sudarsky03}). \texttt{HELIOS} is a Python-based GPU dependent model and it has been used for modeling atmospheres of a variety of substellar objects \citep[e.g.][]{fossati21,rockymalik19,yan22,deline22}. \texttt{TLUSTY} is a stellar atmosphere model written in FORTRAN77. This code has been modified to be applied to substellar atmospheres as well and is called \texttt{COOLTLUSTY} \citep{hubeny03}. \texttt{COOLTLUSTY} also has been widely used for brown dwarfs and exoplanets \citep[e.g.][]{burrows06,spiegel10,spiegel12,lacy19,lacy20}. In this work, we aim to add to develop a new open-source Python based one-dimensional radiative-convective equilibrium model for H-dominated substellar atmospheres -- \texttt{PICASO 3.0} that is both: 1) open source and 2) capable of including disequilibrium chemistry induced by vertical mixing self-consistently within the 1D radiative--convective equilibrium framework. \texttt{PICASO 3.0} has been released publicly as an extension of the already open-source 1D and 3D radiative transfer tool \texttt{PICASO} \citep{batalha19}. It is also accompanied by detailed tutorials exploring all of its uses and limitations. Here, we focus on the description of the numerical techniques used in the code along with benchmarking to previous studies to demonstrate its functionality.

In \S\ref{sec:model}, we discuss the methodology of \texttt{PICASO 3.0}, the new Python atmospheric model. We present the benchmarking of our model with other models in \S\ref{sec:benchmark} followed by recommendations on how to use this model in \S\ref{sec:recommendations}.  We briefly discuss  ongoing and  future improvements in \S\ref{sec:improve} and conclude in \S\ref{sec:summary}.

\section{Model Setup of \texttt{PICASO 3.0}}\label{sec:model}
\begin{figure*}
  \centering
  \includegraphics[width=1\textwidth]{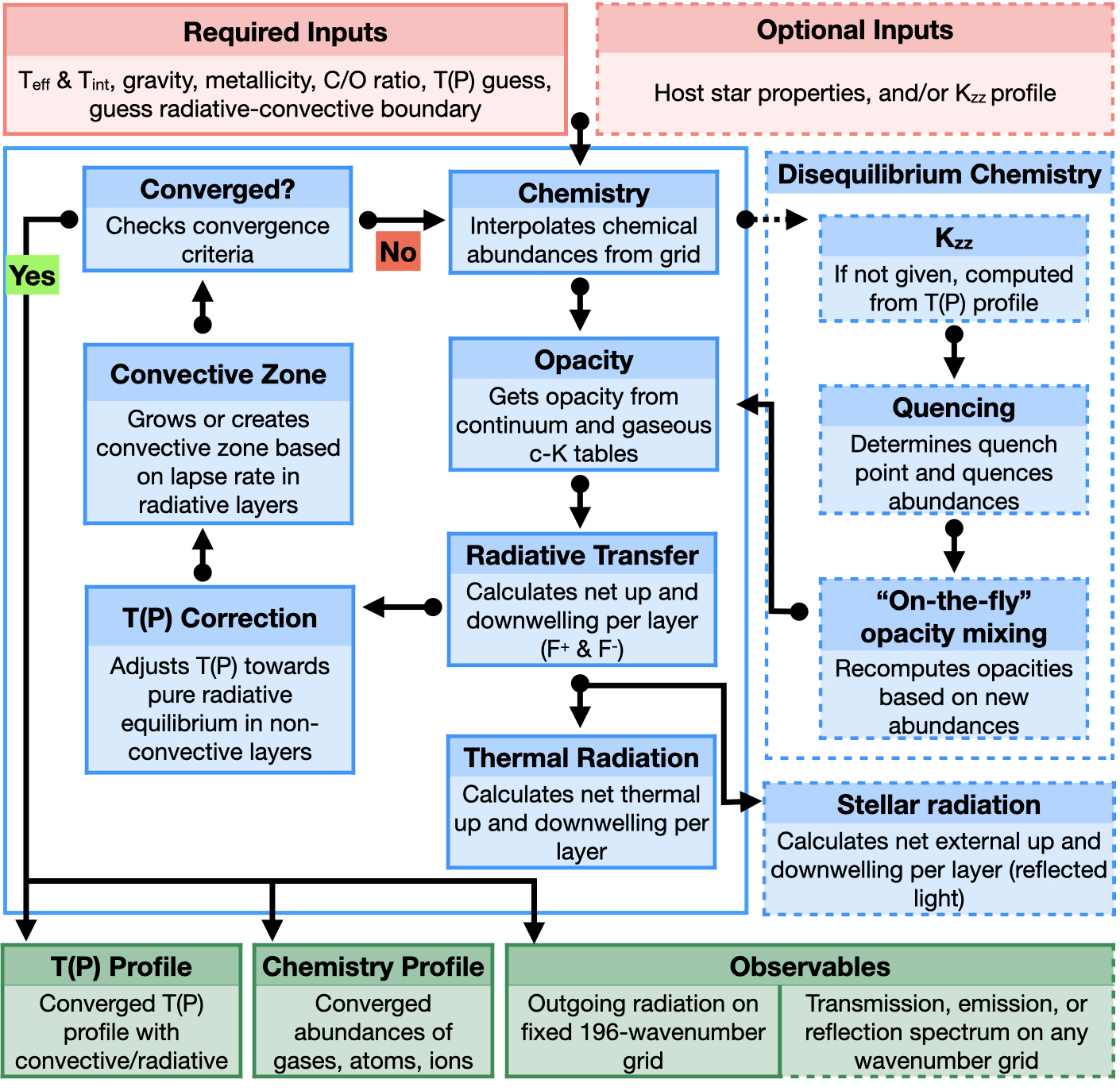}
  \caption{Schematic diagram of the radiative-convective model in \texttt{PICASO 3.0}. Solid outlines indicate required modules, while dashed outlines indicate optional models. }
\label{fig:figschematic}
\end{figure*}

The heritage of our code is the Fortran based \texttt{EGP} sub-stellar atmospheric model. The \texttt{EGP} model has been used for substellar atmospheres including solar system planets -- Titan and Uranus \citep{mckay1989thermal,marley1999thermal}, the L--T--Y brown dwarf sequence \citep[e.g.][]{Marley96,morley14water,morley2012neglected,marley21,karilidi21,zhang21}, cloudy atmospheres \citep[e.g.][]{cushing08,stephens09} and a wide variety of extrasolar planets \citep[e.g.][]{fortney2005comp,fortney2007planetary,fortney08, Marley_2012,fortney20, Morley2015super,Morley2017gj436}  for over two decades. It has also been used as the primary radiative-transfer scheme in the SPARC GCM \citep{Showman2009}. Briefly, this 1D model solves for the self-consistent temperature, chemical and cloud structure of H-dominated atmospheres under the assumption of radiative-convective equilibrium. The code structure of \texttt{PICASO 3.0} is composed of Python classes  \footnote{https://docs.python.org/3/tutorial/classes.html} which generally include multiple Python functions. We refer to these classes and functions as modules here. In general, Python is significantly slower than Fortran. However, with the use of `numba`'s `just-in-time` framework \citep{numba} to create on-the-fly compiled machine code, \texttt{PICASO 3.0} has comparable run times to the original Fortran.  Figure \ref{fig:figschematic} shows a simplified schematic of \texttt{PICASO 3.0}'s workflow. 

First, the user provides an initial set of physical properties for the object to be modeled, shown within the red boxes at the top of Figure \ref{fig:figschematic}. The model supports both non-irradiated (e.g., brown dwarf) and irradiated (e.g., planets) calculations. Therefore, we specify optional inputs with dashed outlines. For example, for modeling a field brown dwarf atmosphere, the user only must specify the {\teff} ({\tint} in case of a planet), gravity, atmospheric metallicity and C/O ratio of the brown dwarf. For modeling an irradiated planet, the user must specify the internal temperature of the planet {\tint}, planet gravity, atmospheric metallicity, C/O ratio, semi-major axis, and host star properties (e.g., stellar temperature, metallicity, and gravity). {\tint} represents the temperature obtained by converting the internal heat flow of the planet via the Stefan-Boltzmann law. Along with these inputs, the user must also specify an initial pressure-temperature profile ($T(P)$) guess, which is divided into plane-parallel logarithmically spaced pressure layers ($\sim$ 60-90). Then, through an iterative process, which we describe in \S \ref{sec:physics}, the model iterates through computing chemistry, opacities, and net upwelling and downwelling fluxes until a radiative-convective equilibrium threshold has been met. During this iteration, the model solves for both radiative and convective parts of the atmosphere and takes into account the possibility of multiple radiative and convective zones. Ultimately, the model produces the final atmospheric state of the object ($T(P)$ and associated chemistry). These can then be used to compute transmission, emission, and/or reflected light spectra used to compare with observations (see  Figure \ref{fig:figschematic}). We discuss these physical and chemical aspects one by one in the following \S \ref{sec:physics}.

\subsection{Physics and Chemistry of Substellar Atmospheres}\label{sec:physics}
\subsubsection{Radiative-Convective Equilibrium}\label{sec:RCE}

The key physical basis of this model is the assumption of radiative-convective equilibrium in substellar atmospheres. The radiative equilibrium represents the physical scenario where within each atmospheric layer the energy emitted must be balanced by the energy absorbed. This means that each atmospheric layer must allow the transfer of the same amount of radiative energy through it. \citet{hubeny17} provides a detailed derivation of the radiative--convective equilibrium criteria.  For completeness, we present some of the important steps here. Radiative equilibrium scenario is represented by:

\begin{equation}\label{eq:eqRCE}
    \int_0^\infty (\chi_{\nu}J_{\nu}-\eta_{\nu})d{\nu} =0
\end{equation}
where $\chi_{\nu}$ and ${\eta_{\nu}}$ are absorption and emission coefficients, respectively and $J_{\nu}$ is the first moment of the intensity field. Assumption of local thermodynamic equilibrium allows Equation \ref{eq:eqRCE} to be rewritten as:

\begin{equation}\label{eq:eqRCE1}
    \int_{\nu} \kappa_{\nu}(J_{\nu}-B_{\nu})d{\nu} =0
\end{equation}
where $\kappa_{\nu}$ is the wavelength dependent total absorption opacity of the layer and $B_{\nu}$ is the local Planck function of the layer. This is the integral form of the radiative equilibrium condition for each atmospheric layer. The integral form of the radiative--equilibrium condition (Equation \ref{eq:eqRCE1}) is applicable and numerically stable throughout the atmosphere. Another method of computing radiative-equilibrium condition, which is used by our model, is referred to as the differential form.  The differential form is numerically stable at optically thicker parts of the atmosphere and might become numerically unstable at parts of the atmosphere which are optically  thin ($\tau <<$ 1). In practice, this instability only appears at parts of the atmosphere with low temperatures and very small optical depths. In our first release of the climate code, we stick to our original methodology of using the differential form, despite the instabilities pointed out by \citet{hubeny17}. In a later update we will explore and implement the integral form as well within our model to make our solutions more accurate at the optically thinner parts of the atmosphere.

The differential form can be derived by using Equation \ref{eq:eqRCE1} along with the second moment of the radiative transfer equation, 
\begin{equation}
\int_{\mu} \mu\dfrac{dI_{\nu}}{d\tau_{\nu}}d\mu = \int_{\mu}\mu(I_{\nu} - S_{\nu})d{\mu}
\end{equation}
 which under local thermodynamic equilibrium leads to the form,

\begin{equation}\label{eq:eqRT}
    \dfrac{dH_{\nu}}{dz}= \kappa_{\nu}(J_{\nu} - B_{\nu})
\end{equation}
where $H_{\nu}$ is the wavelength dependent second moment of the specific intensity field. Comparing Equation \ref{eq:eqRCE1} and \ref{eq:eqRT} leads to the condition,
\begin{equation}\label{eq:eqRT1}
    \int_{\nu}\dfrac{dH_{\nu}}{dz}d{\nu}= 0
\end{equation}
which means that the integral of the second moment of the intensity, {$H_{\nu}$}, must be constant at all layers. Mathematically, this can be written as:
\begin{equation}\label{eq:RCEfin}
    \int_{\nu} H_{\nu}d{\nu} - \dfrac{\sigma_{sb}T_{\rm eff}^4}{4\pi} = 0
\end{equation}
where $\sigma_{sb}$ is the Stefan--Boltzmann constant and \teff\ is the effective temperature of the object. This calculation can be found in the \href{https://github.com/natashabatalha/picaso/blob/caf63752563215e76ec713f65182f7efc367f3fc/picaso/climate.py#L448}{\texttt{t\char`_start}} module in the code. The model tries to achieve this radiative equilibrium in all the atmospheric layers which are stable against convection. 

Given Equation \ref{eq:RCEfin}, which defines the radiative-equilibrium condition, the general procedure of the model is to start with an initial guess of the $T(P)$ profile of the atmosphere. This can be seen in the required inputs box at the top of Figure \ref{fig:figschematic}. Radiative fluxes through all the atmospheric layers are then calculated given a chemical state of the atmosphere. These radiative fluxes are used to compute the wavelength integrated flux carried by each layer $H(z)$. If the atmospheric layers are not in radiative equilibrium, this quantity $H(z)$ would not be equal to the target flux given by $\dfrac{\sigma_{sb}T_{\rm eff}^4}{4\pi}$, and the $T(P)$ solution would be perturbed until the convergence criteria set by Equation \ref{eq:RCEfin} is met. The methodology of the convergence is discussed further in \S\ref{sec:model_convergence}.

\subsubsection{Radiative Transfer}\label{sec:RT}

In order to compute radiative-convective equilibrium with Equation \ref{eq:RCEfin}, radiative fluxes must be calculated for all atmospheric layers. The fundamental equation describing atmospheric radiative transfer is,

\begin{equation}\label{eq:RT1}
\begin{aligned}
    \mu\dfrac{dI_{\nu}(\mu,\tau_{\nu},\phi)}{d\tau_{\nu}}=             &I_{\nu}(\mu,\tau_{\nu},\phi)-S_{\nu}(\mu,\tau_{\nu},\phi) \\
    & - \dfrac{\omega_0}{4\pi}\int_0^{2\pi}\int_{-1}^{1}P_{\nu}(\mu,\mu^{'},\phi,\phi^{'}) \\
    & \cdot I_{\nu}(\mu^{'},\tau_{\nu},\phi^{'})d\mu^{'}d\phi^{'}
\end{aligned}
\end{equation} 
where $\mu = \cos(i)$ where $i$ is the incident angle of the ray, $\phi$ is the azimuthal angle, $I_{\nu}$ is the specific intensity, $\tau_{\nu}$ is the frequency dependent optical depth, $S_{\nu}$ is the frequency dependent source function and $P_{\nu}$ is the scattering phase--function. The first term on the right--hand--side of Equation \ref{eq:RT1} describes the attenuation of the intensity with increasing optical depth. The source function $S_{\nu}$ is used to describe the emission of the atmosphere itself (e.g. thermal emission) or any external radiation source (e.g. incident stellar radiation). The third term in Equation \ref{eq:RT1} represents scattering of the intensity within the atmosphere where the phase--function $P_{\nu}(\mu,\mu^{'},\phi,\phi^{'})$ is the probability that intensity $I_{\nu}(\mu^{'},\tau_{\nu},\phi^{'})$ will be scattered from the direction ($\mu^{'},\phi^{'}$) to the direction ($\mu,\phi$).

In order to solve Equation \ref{eq:RT1}, we follow the two-stream radiative transfer methodology described in \citet{toon1989rapid} for the calculation of radiative fluxes. The specific intensity $I_{\nu}(\mu,\tau_{\nu},\phi)$ can be integrated over the azimuthal angle $\phi$ to calculate the azimuthally integrated intensity $I_{\nu}(\mu,\tau_{\nu})$. The upward and downward fluxes can also be defined by,
\begin{equation}\label{eq:RT2}
\begin{aligned}
    F_{\nu}^{+}= & \int_{0}^{1} {\mu}I^{+}_{\nu}(\mu,\tau_{\nu})d\mu \\
    F_{\nu}^{-}= & \int_{0}^{1} {\mu}I^{-}_{\nu}(\mu,\tau_{\nu})d\mu
\end{aligned}
\end{equation} 
where $I^{+}_{\nu}(\mu,\tau_{\nu})$ is the azimuthally integrated intensity for upward values of $\mu$ and $I^{-}_{\nu}(\mu,\tau_{\nu})$ is the azimuthally integrated intensity for downward values of $\mu$. Equation \ref{eq:RT1} can be integrated to produce two separate coupled equations in terms of the upward and downward fluxes instead of the $\mu$ and $\phi$ dependent intensities. These equations for the upward and downward fluxes are\footnote{there is a typo in Eqn. 12 of \citet{toon1989rapid} which incorrectly swaps in $S^{+}_{\nu}$ for $S^{-}_{\nu}$ in the negative partial flux},

\begin{equation}\label{eq:RT3}
\begin{aligned}
    \dfrac{{\partial}F_{\nu}^{+}}{\partial\tau_{\nu}}= & \gamma_1F_{\nu}^{+} - \gamma_2F_{\nu}^{-} -S^{+}_{\nu} \\
    \dfrac{{\partial}F_{\nu}^{-}}{\partial\tau_{\nu}}= & \gamma_2F_{\nu}^{+} - \gamma_1F_{\nu}^{-} +S^{-}_{\nu}
\end{aligned}
\end{equation}
where $\gamma_1$ and $\gamma_2$ are functions of the scattering properties of the medium and $S^{+}_\nu$ and $S^{-}_\nu$ are modified versions of the source function. The radiative transfer calculation within our model is divided into two distinct parts: 1) the transfer of thermal emission through the atmosphere and 2) the transfer of the reflected external (e.g. stellar) radiation throughout the atmosphere. Equation \ref{eq:RT3} is applicable to both of these components but the functions $\gamma_1$, $\gamma_2$, $S^{+}_\nu$, and $S^{-}_\nu$ are defined differently for each. A brief description of the radiative transfer of thermal radiation in our model is provided below followed by a discussion of the radiative transfer of external stellar radiation.

We use the two-stream source function technique described in \citet{toon1989rapid} to calculate the thermal upward and downward fluxes in each layer. The hemispheric mean approximation approach is used in this method where $\gamma_1$ is given by 2-$\omega_0(1+g)$ and $\gamma_2$ is $\omega_0(1+g)$. $\omega_0$ and $g$ are the single scattering albedo and the scattering asymmetry parameter of the atmospheric layer, respectively. This hemispheric mean approximation was used to obtain functional forms for the source functions $S^{+}_\nu$ and $S^{-}_\nu$ in \citet{toon1989rapid} (see Table 3 and Equations 53 \& 54 in \citet{toon1989rapid}). With these source functions, the upward azimuthally averaged intensities at the top ($I^{+}_{n}(0,\mu)$) and bottom ($I^{+}_{n}(\tau,\mu)$) of an atmospheric layer with optical depth $\tau$ can be written as,

\begin{equation}\label{eq:RT4}
\begin{aligned}
    I^{+}_{n}(0,\mu) = & I^{+}_{n}(\tau,\mu)e^{-\tau/\mu} + \dfrac{G}{\lambda\mu-1}(e^{\lambda\tau}e^{-\tau/\mu}-1) \\
                       & + \dfrac{H}{\lambda\mu+1}(1-e^{-\lambda\tau}e^{-\tau/\mu}) +\alpha_1(1-e^{-\tau/\mu}) \\
                       & + \alpha_2(\mu-(\tau+\mu)e^{-\tau/\mu})
\end{aligned}
\end{equation} 
This calculation can be found in the \href{https://github.com/natashabatalha/picaso/blob/caf63752563215e76ec713f65182f7efc367f3fc/picaso/fluxes.py#L1750}{\texttt{get\char`_thermal\char`_1d\char`_gfluxi}} module. Similarly the downward azimuthally averaged intensities at the top ($I^{-}_{n}(0,-\mu)$) and bottom ($I^{-}_{n}(\tau,-\mu)$) of the same atmospheric layer can be written as,

\begin{equation}\label{eq:RT5}
\begin{aligned}
    I^{-}_{n}(\tau,-\mu) = & I^{-}_{n}(0,-\mu)e^{-\tau/\mu} + \dfrac{K}{\lambda\mu-1}(e^{-\tau/\mu}-e^{-\lambda\tau}) \\
                       & + \dfrac{J}{\lambda\mu+1}(e^{\lambda\tau}-e^{-\tau/\mu}) +\sigma_1(1-e^{-\tau/\mu}) \\
                       & + \sigma_2({\mu}e^{-\tau/\mu}+\tau-\mu)
\end{aligned}
\end{equation} 
and can also be found in the \href{https://github.com/natashabatalha/picaso/blob/caf63752563215e76ec713f65182f7efc367f3fc/picaso/fluxes.py#L1736}{\texttt{get\char`_thermal\char`_1d\char`_gfluxi}} module. The functions $G$, $H$, $K$, $I$, $\alpha_1$, $\alpha_2$, $\sigma_1$, and $\sigma_2$ have been computed for the hemispheric-mean approximation in \citet{toon1989rapid} and are used in our calculation as well. Solving equations \ref{eq:RT4} and \ref{eq:RT5} also require boundary conditions for the diffuse flux at the top and bottom of the atmosphere. These boundary conditions are set using the thermal blackbody intensities at the top and bottom of the atmosphere using,

\begin{equation}\label{eq:RT6}
\begin{aligned}
    B_{\rm top}= & (1.-e^{-\tau^{'}/\mu_1})B({ T_{\rm top}}) \\
    B_{\rm bot} = & B({ T_{\rm bot}})+\mu_1\dfrac{B({ T_{\rm bot}})-B({ T_{\rm bot-1}})}{\tau}\\
\end{aligned}
\end{equation} 

where $B(T)$ represents the blackbody function, $T_{\rm top}$ is the temperature of the top most atmospheric layer, $T_{\rm bot}$ is the temperature of the bottom most atmospheric layer, and $\tau$ is the optical depth of the bottom most atmospheric layer. $\mu_1$ is assumed to be 0.5 due to the hemispheric-mean approximation. Note that the bottom boundary condition is valid only for gas giant atmospheres where the highest pressure grid point corresponds to the end of the user-defined grid and  does not correspond to a ``surface''. \texttt{PICASO 3.0} does have the option to swap boundary conditions that are pertinent to the hard surfaces needed for terrestrial atmospheres, but it is not relevant for this work on gas giants and as such we do not discuss it here.   $\tau^{'}$ is given by,

\begin{equation}\label{eq:RT7}
    \tau^{'}= \tau_{top}\dfrac{{ P_{\rm top}}}{({ P_{\rm top+1}}-{ P_{\rm top}})}
\end{equation}

This expression along with the top boundary condition in Equation \ref{eq:RT6} captures the downward thermal flux arising from the part of the atmosphere which has pressure less than the minimum pressure in the used atmospheric pressure grid. This formulation prevents arbitrary artificial cooling of the top-most atmospheric layer.

Equations \ref{eq:RT4} and \ref{eq:RT5} can be used to calculate the incidence angle ($\mu$) dependent upward and downward -- $I_n^{+}(\mu)$ and $I_n^{-}(\mu)$  intensity field in each atmospheric layer. But in order to use these radiative intensities ultimately for the convergence criteria in Equation \ref{eq:RCEfin}, they need to be integrated with Equation \ref{eq:RT2} to calculate the direction independent upward and downward fluxes -- $F^{+}_{\nu}$ and $F^{-}_{\nu}$. In order to compute these disk-averaged, layer fluxes, the upward and downward thermal intensity for each atmospheric layer is calculated using Equation \ref{eq:RT4} and \ref{eq:RT5} at five incident angles. The cosine of these incident angles ($\mu$) are determined using the Gaussian quadrature method with 5 points\citep{abramowitz}. The choice of 5 Gauss points is the default setting in the code but other choices are also available in the \href{https://github.com/natashabatalha/picaso/blob/caf63752563215e76ec713f65182f7efc367f3fc/picaso/disco.py#L46}{\texttt{get\char`_angles\char`_1d}} module. Table \ref{table:tab1} shows the five default values of $\mu$ and the corresponding Gauss weights used for this Gaussian quadrature integration. The intensities at different incident angles are used to compute the integral in Equation \ref{eq:RT2} with the Gaussian-quadrature integration technique using the weights in Table \ref{table:tab1}. The module \href{https://github.com/natashabatalha/picaso/blob/caf63752563215e76ec713f65182f7efc367f3fc/picaso/fluxes.py#L1551}{\texttt{get\char`_thermal\char`_1d\char`_gfluxi}} uses the formulation described above to calculate the wavelength dependent upward and downward thermal fluxes at the edges of each atmospheric layer.

\begin{table}[h!]
\centering
\begin{tabular}{||c c c||} 
 \hline
 $\theta$(deg) & $\mu=cos(\theta)$ & Gauss Weight \\ [0.5ex] 
 \hline\hline
 84.345 & 0.09853 & 0.015747 \\ 
 72.270 & 0.30453 & 0.073908 \\
 55.804 & 0.56202 & 0.146386 \\
 36.680 & 0.80198 & 0.167174 \\
 16.221 & 0.96019 & 0.096781 \\ [1ex] 
 \hline
\end{tabular}
\caption{Gauss points and weights used for Gaussian quadrature integration of thermal flux over different angles.}
\label{table:tab1}
\end{table}

For the radiative transfer of the reflected stellar light, we use the quadrature approximation. The calculation is performed for a single incidence angle of the stellar radiation beam (60$^{\circ},\mu_0$=0.5)  unlike the 5 Gauss point method of the thermal counterpart. Under the quadrature approximation, $\gamma_1$  is $0.5\sqrt{3}(2-\omega_0(1+g))$, $\gamma_2$ is $\omega_0\sqrt{3}(1-g)/2$, $\gamma_3$ is ($1-\sqrt{3}g\mu_0$)/2, and $\gamma_4$ is 1-$\gamma_3$. These functions can be found in the \href{https://github.com/natashabatalha/picaso/blob/caf63752563215e76ec713f65182f7efc367f3fc/picaso/fluxes.py#L1219}{\texttt{get\char`_reflected\char`_1d\char`_gfluxv}} module. The source functions $S_{\nu}^{+}$ and $S_{\nu}^{-}$ for this component are,

\begin{equation}\label{eq:RT8}
\begin{aligned}
    S_{\nu}^{+}= & \gamma_3{\pi}F_s\omega_0e^{-(\tau_c+\tau)/\mu_0} \\
    S_{\nu}^{-}= & \gamma_4{\pi}F_s\omega_0e^{-(\tau_c+\tau)/\mu_0}\\
\end{aligned}
\end{equation} 

where $\tau_c$ is the cumulative optical depth of all the atmospheric layers above the layer of calculation and $\tau$ is the optical depth of the layer itself. $F_s$ here represents the stellar flux incident on the top of the atmosphere. $F_s$ is interpolated from the \texttt{PHOENIX} grid of models \citep{allard12} available as a part of the \texttt{PySynPhot} \citep{pysynphot13} package. Using these functions, Equation \ref{eq:RT3} can be solved for the upward and downward fluxes for the $n$'th atmospheric layer as has been shown in \citet{toon1989rapid},

\begin{equation}\label{eq:RT9}
\begin{aligned}
    F_{\nu,n}^{+}= & k_{1n}e^{\lambda_n\tau_n} + \Gamma_n{k_{2n}}e^{-\lambda_n\tau_n} + C_n^{+} \\
    F_{\nu,n}^{-}= & \Gamma_{n}k_{1n}e^{\lambda_n\tau_n} + {k_{2n}}e^{-\lambda_n\tau_n} + C_n^{-}\\
\end{aligned}
\end{equation} 

where the quantities $\Gamma_n$, $k_{1n}$, $k_{2n}$, $\lambda_n$, $C_n^{+}$, and $C_n^{-}$ are defined in \citet{toon1989rapid} (Equation 21, 22, 23, and 24). Equation \ref{eq:RT9} is solved in the \href{https://github.com/natashabatalha/picaso/blob/caf63752563215e76ec713f65182f7efc367f3fc/picaso/fluxes.py#L1288}{\texttt{get\char`_reflected\char`_1d\char`_gfluxv}} module. However, an additional term needs to be added to the downward fluxes solved from Equation \ref{eq:RT9} in this formulation. This term is,

\begin{equation}\label{eq:RT10}
\begin{aligned}
    F_{\nu,n}^{-}= F_{\nu,n}^{-} + \mu_{0}{F_s}e^{-\tau_c/\mu_0} \\
\end{aligned}
\end{equation} 

Like the thermal component, complete solutions of Equation \ref{eq:RT9} also require boundary conditions on the diffuse flux at the top and bottom of the atmosphere. We use the boundary conditions,

\begin{equation}\label{eq:RT11}
\begin{aligned}
    F_{\rm top,st}= & 0 \\
    F_{\rm bot,st} = & {R_s}\mu_{0}F_{s}e^{-\tau_{\rm c,bot}/\mu_0}\\
\end{aligned}
\end{equation} 

where $F_{\rm top,st}$ and $F_{\rm bot,st}$ are the diffuse flux at the top and bottom of the atmosphere, $R_s$ denotes the reflectivity of the bottom surface and $\tau_{c,bot}$ is the cumulative optical depth of the deepest layer. The reflectivity of the bottom surface is typically assumed to be 1\% in the model, but users should check that the bottom surface is not optically thin. This is further discussed in the \S\ref{sec:recommendations}. This formulation is used by the \href{https://github.com/natashabatalha/picaso/blob/caf63752563215e76ec713f65182f7efc367f3fc/picaso/fluxes.py#L1123}{\texttt{get\char`_reflected\char`_1d\char`_gfluxv}} module to calculate the wavelength dependent upward and downward reflected stellar light fluxes at the edges of each atmospheric layer. But the calculation of both the thermal and reflected stellar light fluxes  following the procedure described above requires another  important parameter -- the optical depth $\tau$. The implemented $\tau$ calculation procedure in this model is described below.

%Another very important aspect of Equation \ref{eq:RT1} is the optical depth $\tau$ which is required to calculate both the thermal and stellar incident fluxes by these equations. 

Generally, two approaches are used for calculating the optical depth $\tau$ -- 1) the line--by--line approach (e.g. \citet{gandhi17},\citet{burrows08}) or 2) the correlated-k approach (e.g. \citet{marley1999thermal}, \citet{malik17}, \citet{fortney08}). This model uses a correlated-k opacity approach for computationally efficient inclusion of gaseous opacities. In the correlated-k approach, each wavelength bin is represented by 8 distinct Gaussian quadrature points ($g_i$), each with its associated weight ($w_i$). Table \ref{table:tab2} shows the Gauss points and the Gauss weights used for the 8 point correlated-k approach in the model. Each of these Gauss-points also have an optical depth $\tau_i$ associated with them. These ``Gauss" points should not be confused with the ``Gauss" points used to calculate the disk-integrated fluxes. This means that the radiative transfer equations for each wavelength bin must be computed 8 times, once for each value of optical depth $\tau_i$ corresponding to the $i$'th Gauss point. This calculation is done by calling \href{https://github.com/natashabatalha/picaso/blob/caf63752563215e76ec713f65182f7efc367f3fc/picaso/fluxes.py#L1551}{\texttt{get\char`_thermal\char`_1d\char`_gfluxi}} and \href{https://github.com/natashabatalha/picaso/blob/caf63752563215e76ec713f65182f7efc367f3fc/picaso/fluxes.py#L1123}{\texttt{get\char`_reflected\char`_1d\char`_gfluxv}} modules 8 times by the \href{https://github.com/natashabatalha/picaso/blob/caf63752563215e76ec713f65182f7efc367f3fc/picaso/climate.py#L1122}{\texttt{climate}} module. The 8 thermal and radiative upward and downward fluxes are then added together following,

\begin{equation}\label{eq:RT12}
    F^{\pm}_n(\lambda) = \sum_{i=1}^{i=8} w_iF^{\pm}_{i,n}(\lambda,\tau_{i,n})
\end{equation}

where $F^{\pm}_{i,n}$ is the thermal/reflected light flux at the $n$'th atmospheric layer calculated with the optical depth $\tau_{i,n}$ corresponding to the $i$'th Gauss point. Our calculation of correlated-k opacities is described in more details in \S\ref{sec:opa}. This summation with Equation \ref{eq:RT12} within the \texttt{climate} module ultimately produces thermal/reflected wavelength dependent upward and downward fluxes in each atmospheric layer. These fluxes can now be used to calculate the convergence criteria in Equation \ref{eq:RCEfin}. The net thermal flux for the $n$'th layer can be simply obtained with,
\begin{equation}\label{eq:RT13}
\begin{aligned}
    F^n_{\rm thermal,net} = & \int_{\lambda}( F^+_n(\lambda) -F^-_n(\lambda))d\lambda \\
    F^n_{\rm stellar,net} = & \int_{\lambda}( F^+_{n,s}(\lambda) -F^-_{n,s}(\lambda))d\lambda\\
\end{aligned}
\end{equation}
where $F^+_n(\lambda)$ represents the upward thermal flux from the $n$'th layer within the wavelength bin between $\lambda$ and $\lambda$+d$\lambda$ and $F^-_n(\lambda)$ is the downward thermal flux from the same layer within the same wavelength bin. $F^{\pm}_{n,s}$ also represent the same quantity for the reflected light component. The total net radiative flux in the $n$'th atmospheric layer can be calculated using,

\begin{equation}\label{eq:net}
    F^n_{\rm net} = r_{\rm th}F^n_{\rm thermal,net} + r_{\rm st}F^n_{\rm stellar,net}
\end{equation}

where r$_{\rm th}$ is the contribution factor of the thermal radiation to the net flux and is generally fixed at 1 and r$_{\rm st}$ is the contribution factor of the stellar radiation to the net flux. Non-zero values of r$_{\rm st}$ is only relevant when the external irradiation on the atmosphere is non-zero. In the scenario when a user is computing a planet-wide average $T(P)$ profile, the stellar irradiation is contributing to 50\% (one hemisphere) of the planet and as a result $r_{\rm st}=0.5$. If instead the goal is to compute a night-side average atmospheric state, $r_{\rm st}$ is set to be 0. On the other extreme, to compute the day-side atmospheric state of a tidally locked planet $r_{\rm st}$ should be set at 1.

This full net flux, $F^n_{\rm net}$, is same as $4\pi\int_{\nu}H_{\nu}d\nu$ in Equation \ref{eq:RCEfin}. Therefore, $F^n_{\rm net}$ can be used to check if the convergence criteria (Equation \ref{eq:RCEfin}) is satisfied at all the radiative layers of the atmosphere. This check is done in the ``Converged?" box of the flow chart shown in Figure \ref{fig:figschematic}. Now that we have described the radiative transfer within our model, we will discuss other blocks of the model flow chart shown in Figure \ref{fig:figschematic} starting with atmospheric chemistry.

\begin{table}[h!]
\centering
\begin{tabular}{||c | c||} 
 \hline
 Gauss Points ($g_i$) & Gauss Weights ($w_i$)\\ [0.5ex] 
 \hline\hline
 0.065960251992824 & 0.165231051440291 \\ 
 \hline
 0.313509004297193 & 0.309768948559709\\
 \hline
 0.636490995702807 & 0.309768948559709\\
 \hline
 0.884039748007175 & 0.165231051440291\\
 \hline
 0.953471592210149 & 0.008696371128436358\\
 \hline
 0.966500473910379 & 0.01630362887156367\\
 \hline
 0.983499526089621 & 0.01630362887156367\\
 \hline
 0.996528407789851 & 0.008696371128436358\\[1ex] 
 \hline
\end{tabular}
\caption{Gauss points and weights used for calculation of the correlated-k opacities}
\label{table:tab2}
\end{table}

\subsubsection{Equilibrium Chemistry}\label{sec:chem}

As shown in Figure \ref{fig:figschematic}, in order to compute the radiative fluxes for checking the convergence criteria, as outlined in the previous \S \ref{sec:RT}, we first need a method to provide the chemical state of the atmosphere. This is required because the chemistry dictates the optical depth ($\tau$) required for the radiative transfer calculations. In the simplest case, we determine the chemical state of the atmosphere using the $T(P)$ profile of the atmosphere, atmospheric metallicity, and C/O ratio assuming chemical equilibrium. The atmospheric metallicity (M/H) is defined as the ratio of the abundances of all heavy elements to Hydrogen abundance in the atmosphere.  Our climate model uses a pre-calculated grid
of molecular abundances on a pressure vs. temperature vs. [M/H] vs. C/O grid. This chemistry grid is computed using the thermochemical equilibrium models presented in \citet{gordon1994computer,lodders99,lodders02,visscher06,channon10} and using protosolar elemental abundances from \citet{Lodders10}. The grid includes 73 temperature points between 75 K and 4000 K and 20 pressure points logarithmically spaced between 10$^{-6}$ and 3000 bars. This corresponds to a pre-computed grid with 1460 grid points. The molecular abundances included in this grid are H$_{2}$, H, H$^{+}$, H$^{−}$, H$_2{^{-}}$, H$_2^{+}$, H$_3^{+}$, He, H$_2$O, CH$_4$, CO, CO$_2$, OCS, HCN, C$_2$H$_2$, C$_2$H$_4$, C$_2$H$_6$, NH$_3$, N$_2$, PH$_3$, H$_2$S, SiO, TiO, VO, Fe, FeH, MgH, CrH, Na, K, Rb, Cs, Li, LiOH, LiH, LiCl, and e-. The metallicities included in the grid are: [M/H] (relative to Solar)= -1, -0.75, -0.5, -0.3, -0.25, 0.0, 0.5, 0.7, 1.0, 1.5, 1.7 and 2. The C/O ratios included in the grid are:  C/O (relative to Solar) = 0.25, 0.5, 1.0, 1.5, 2, and 2.5. In this way, Solar values are [M/H]=0 and C/O=1. In order to change the C/O ratio for a given atmospheric metallicity, both the elemental abundances ratios -- C/H and O/H are increased/decreased slightly while maintaining a constant (C+O)/H. The version of \texttt{PICASO 3.0}, described in this work, does not allow to change other elemental ratios ( e.g. S/H or N/H ). Future releases will include this flexibility.

\begin{figure*}
  \centering
  \includegraphics[width=1\textwidth]{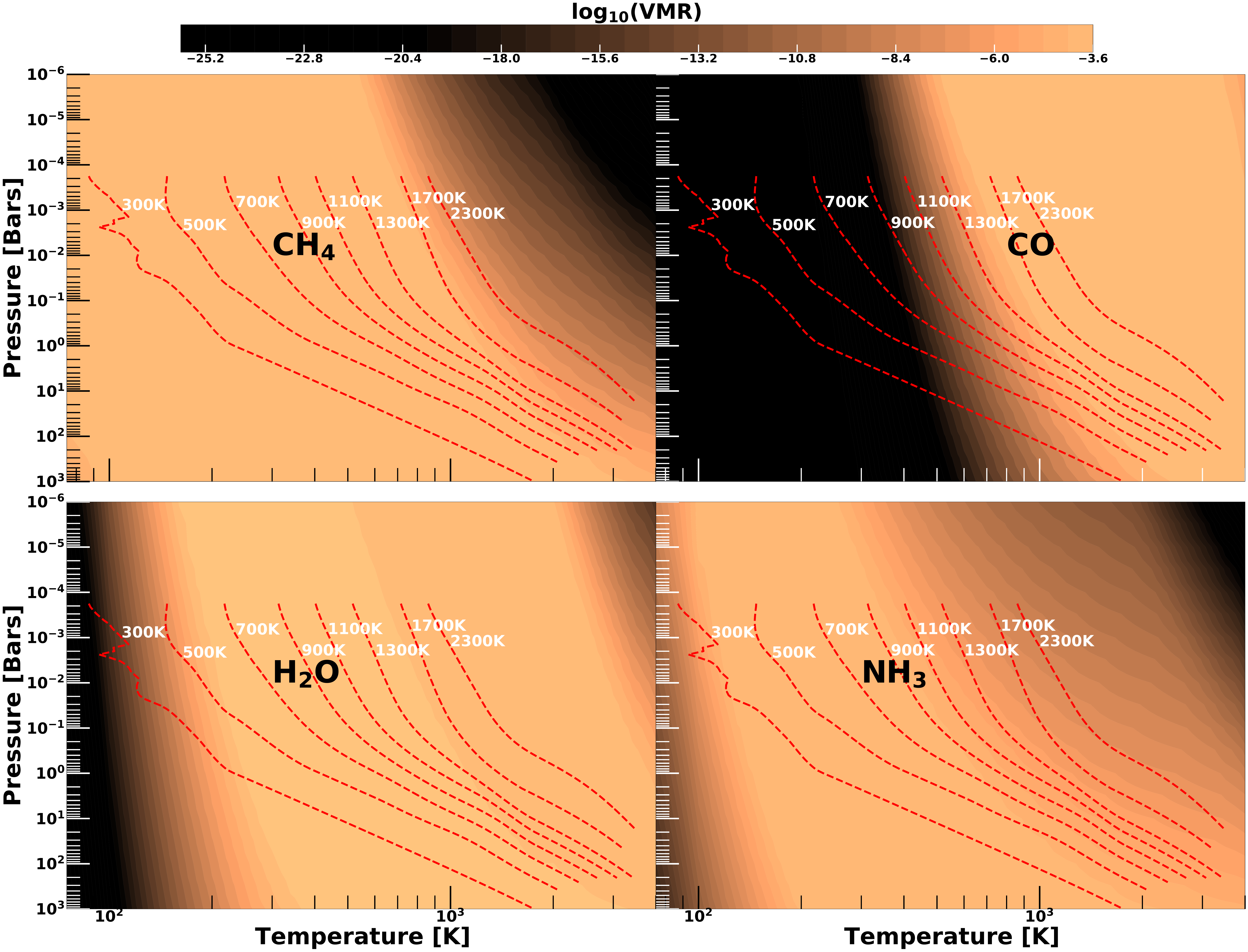}
  \caption{{\bf Top left panel} shows the abundance map of {\meth} in the pressure-temperature chemistry grid used in the model with equilibrium chemistry described in the paper. Dashed red lines marked on the abundance map depict different converged $T(P)$ profiles of brown dwarfs from \citet{marley21} with {\teff} between 300 K and 2300 K and log(g)=5. This grid is used for interpolation of chemical abundances during iteration of the atmospheric state in the code. The same for {\co}, {\water}, and {\amon} is shown in the {\bf top right}, {\bf bottom left} and {\bf bottom right} panel. The abundances shown here are calculated for solar metallicity and C/O ratio. }
\label{fig:figabun}
\end{figure*}

Figure \ref{fig:figabun} shows the pre-computed volume mixing ratios of four major atmospheric gases -- {\meth}, {\co}, {\water}, and {\amon} for a solar mixture of chemical elements. Converged brown dwarf $T(P)$ profiles with log(g)=5 and multiple {\teff} values between 300 K and 2300 K are also overplotted in the pressure-temperature space to depict parts of the parameter space relevant for objects with different {\teff}. There are a few important features that strongly influence the climates of sub-stellar atmospheres that are worth noting. First, the sharp drop in {\water} vapor abundance at $T\lessapprox$200~K  which can be seen in Figure \ref{fig:figabun} lower left panel. This  drop in {\water} vapor abundance is caused by the condensation of {\water} into cloud particles. A similar condensation effect can also be seen in the case of {\amon} at $T\le$100 K. Condensation induced changes in the gaseous abundances are included in the pre-computed grid of gaseous abundances.  Second, the most abundant carbon bearing gas in the atmosphere also changes from \co\ to \meth\ between temperatures of $\sim$ 1000 and 1200 K. Figure \ref{fig:figabun} top right panel shows that \co\ is the major carbon bearing gas above $\sim$ 1200 K but at temperatures cooler than 1200 K, \co\ abundance decreases  rapidly and \meth\ abundance rises. This change can be understood by the net chemical reaction responsible for inter-conversion between \co\ and \meth\ under chemical equilibrium, 

\begin{equation}
    {\rm CH_4 + H_2O \leftrightarrow CO + 3H_2} \\
\end{equation}
This reaction is favored in the forward direction for higher temperatures than $\sim$ 1200 K and as result \co\ is more dominant for such temperatures. The reverse reaction dominates for temperatures lower than $\sim$ 1200 K and \meth\ becomes the dominant C-bearing gas there. Many other similar interesting trends has been explored in the literature for several gases \citep[e.g.][]{lodders02,visscher06,visscher10,moses05,moses11,moses13}. These trends can be directly explored from the available chemical grid within the model but we move on to another challenge of using a pre-computed chemistry grid -- limited number of grid points.

Due to the finite number of grid points (1460 points), chemical abundances cannot be calculated exactly at any pressure--temperature point of interest. Therefore, each time the chemistry routine \href{https://github.com/natashabatalha/picaso/blob/caf63752563215e76ec713f65182f7efc367f3fc/picaso/justdoit.py#L1527}{\texttt{premix\char`_atmosphere}} is called, the climate model uses an interpolation scheme to compute abundances at each pressure-temperature layer point. This interpolation is performed by the module \href{https://github.com/natashabatalha/picaso/blob/caf63752563215e76ec713f65182f7efc367f3fc/picaso/justdoit.py#L2247}{\texttt{chem\char`_interp}}. The 2D interpolation scheme used relies on finding the four surrounding grid points namely $T_{\rm low}(P_{\rm low}$), $T_{\rm low}(P_{\rm high}$), $T_{\rm high}(P_{\rm low}$), and $T_{\rm high}(P_{\rm high}$). Then the abundances in these surrounding points represented by $\xi^i_{low,low}$, $\xi^i_{low,high}$, $\xi^i_{high,low}$, and $\xi^i_{high,high}$ are used to interpolate the abundance of each species in the $T(P)$ point. This interpolation is done using,

\begin{align}
    \ln\xi^i(T,P) = (1-t_{\rm int})(1-p_{\rm int})\ln\xi{^i_{low,low}} +  & \\
                    t_{\rm int}(1-p_{\rm int})\ln\xi{^i_{high,low}} +t_{\rm int}p_{\rm int}\ln\xi{^i_{high,high}} + \nonumber &\\
                    (1-t_{\rm int})p_{\rm int}\ln\xi{^i_{low,high}} \nonumber
\end{align}

where $t_{\rm int}$ and  $p_{\rm int}$ are given by,

\begin{align}
    t_{\rm int} = \dfrac{1/T-{1}/{T_{low}}}{{1}/{T_{high}}-{1}/{T_{low}}} \\
    p_{\rm int} = \dfrac{lnP-lnP_{low}}{lnP_{high}-lnP_{low}}
\end{align}
In the case where the $T(P)$ moves  outside the edges of this chemistry grid,  a linear interpolation scheme is instead adopted within the range 50-5200~K. This allows some flexibility in the iteration scheme. However, beyond these values, the model is not valid. Note, choices in interpolation routines cause discrepancies in resulting abundances. For example, only using two nearest-neighbors causes instabilities in chemical profiles. The interpolation chosen here was tested in multiple chemical regimes to ensure stability. After computing the atmospheric chemistry, the next step in Figure \ref{fig:figschematic} is the computation of the atmospheric opacities, which we describe next.

\subsubsection{Pre-mixed Opacities}\label{sec:opa}

The chemical structure of the atmosphere dictates the optical depth ($\tau(\lambda)$) of each atmospheric layer which is necessary for solving the radiative transfer equations described in \S\ref{sec:RT}. This calculation of atmospheric optical depths occurs in the ``Opacity" block of the model flow chart in Figure \ref{fig:figschematic}. As previously stated, we use the correlated-k approach to handle the molecular opacities in this model. An alternative approach is the line--by--line method which is more precise but comparatively speaking, more computationally expensive than the correlated-k approach at low resolution, and over large wavelength ranges. For completeness, we describe the general methodology of our approach here. However, the data were originally computed by \citet{lupu_roxana_2021_cks}, available on Zenodo, and detailed in \citet{marley21}. We include a table of the main reference data used to compute the pre-mixed opacities in Table \ref{tab:my_label}.  

\begin{table*}
    \centering
    \begin{tabular}{c|c}
         C$_2$H$_2$ &  \citet{hitran2012}\\
         C$_2$H$_4$ & \citet{hitran2012}\\
         C$_2$H$_6$ & \citet{hitran2012} \\
         CH$_4$ &  \citet{yurchenko13vibrational, yurchenko_2014}\\
         CO &  \citet{HITEMP2010,HITRAN2016,li15rovibrational}\\
         CO$_2$ &  \citet{HUANG2014reliable}\\
         CrH &  \citet{Burrows02_CrH}\\
         Fe &  \citet{Ryabchikova2015,oBrian1991Fe,Fuhr1988Fe, Bard1991Fe,Bard1994Fe} \\
         FeH &  \citet{Dulick2003FeH, Hargreaves2010FeH} \\
         H$_2$ & \citet{HITRAN2016} \\
         H$_3^+$ &  \citet{Mizus2017H3p}\\
         H$_2$--H$_2$ & \citet{Saumon12} with added overtone from \citet{Lenzuni1991h2h2} Table 8\\
         H$_2$--He &  \citet{Saumon12} \\
         H$_2$--N$_2$ &  \citet{Saumon12} \\
         H$_2$--CH$_4$ &  \citet{Saumon12} \\
         H$_2^-$  &  \citet{bell1980free}\\
         H$^-$ bf &  \citet{John1988H}\\
         H$^-$ ff &  \citet{Bell1987Hff}\\
         H$_2$O &  \citet{Polyansky2018H2O}\\
         H$_2$S &  \citet{azzam16exomol}\\
         HCN &  \citet{Harris2006hcn,Barber2014HCN,hitran2020}\\
         LiCl &  \citet{Bittner2018Lis}\\
         LiF &  \citet{Bittner2018Lis}\\
         LiH & \citet{Coppola2011LiH} \\
         MgH & \citet{Yadin2012MgH,GharibNezhad2013MgH} assembled in \citet{GharibNezhad2021}\\
         N$_2$ &  \citet{hitran2012}\\
         NH$_3$ &  \citet{yurchenko11vibrationally,Wilzewski16} \\
         OCS &  \citet{HITRAN2016}\\
         PH$_3$ & \citet{sousa14exomol} \\
         SiO &  \citet{Barton2013SiO} \\
         TiO & \citet{McKemmish2019TiO} assembled in \citet{GharibNezhad2021}\\
         VO &   \citet{McKemmish16} assembled in \citet{GharibNezhad2021}\\
         Li,Na,K &  \citet{Ryabchikova2015,Allard2007AA, Allard2007EPJD,Allard2016, Allard2019}\\
         Rb,Cs & \\
         
    \end{tabular}
    \caption{Data used to compute correlated-K opacities. Correlated-K opacities are available at \citet{lupu_roxana_2021_cks} and detailed in \citet{marley21}.}
    \label{tab:my_label}
\end{table*}

\begin{figure*}
  \centering
  \includegraphics[width=1\textwidth]{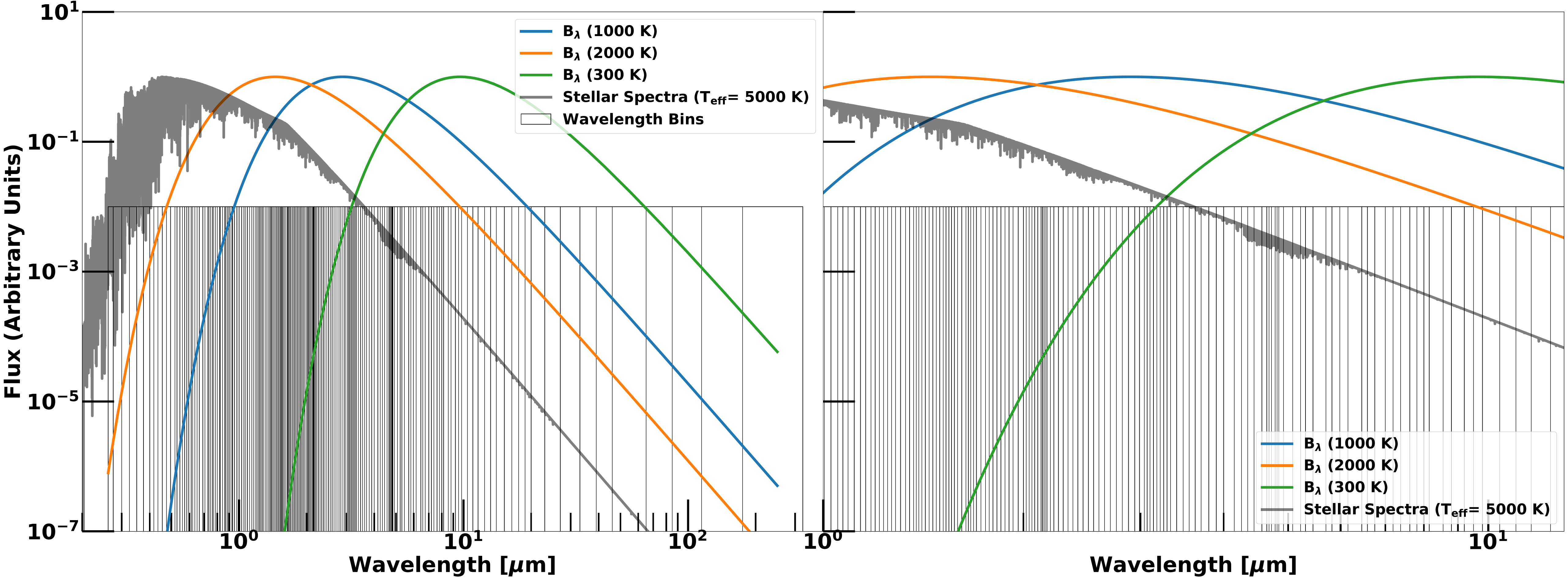}
  \caption{{\bf Left panel} shows the correlated-k opacity wavelength bins used in the radiative transfer calculations in our atmospheric model. The blackbody emission curves of temperatures of 1000 K, 2000 K and 300 K are also shown here. {\bf Right panel} shows a zoomed version of the left panel with the wavelengths between 0.8-13 $\mu$m. The stellar spectrum from a solar metallicity PHOENIX model for a star with {\teff}= 5000 K is also shown in both the panels.}
\label{fig:figwbin}
\end{figure*}

\begin{figure*}
  \centering
  \includegraphics[width=1\textwidth]{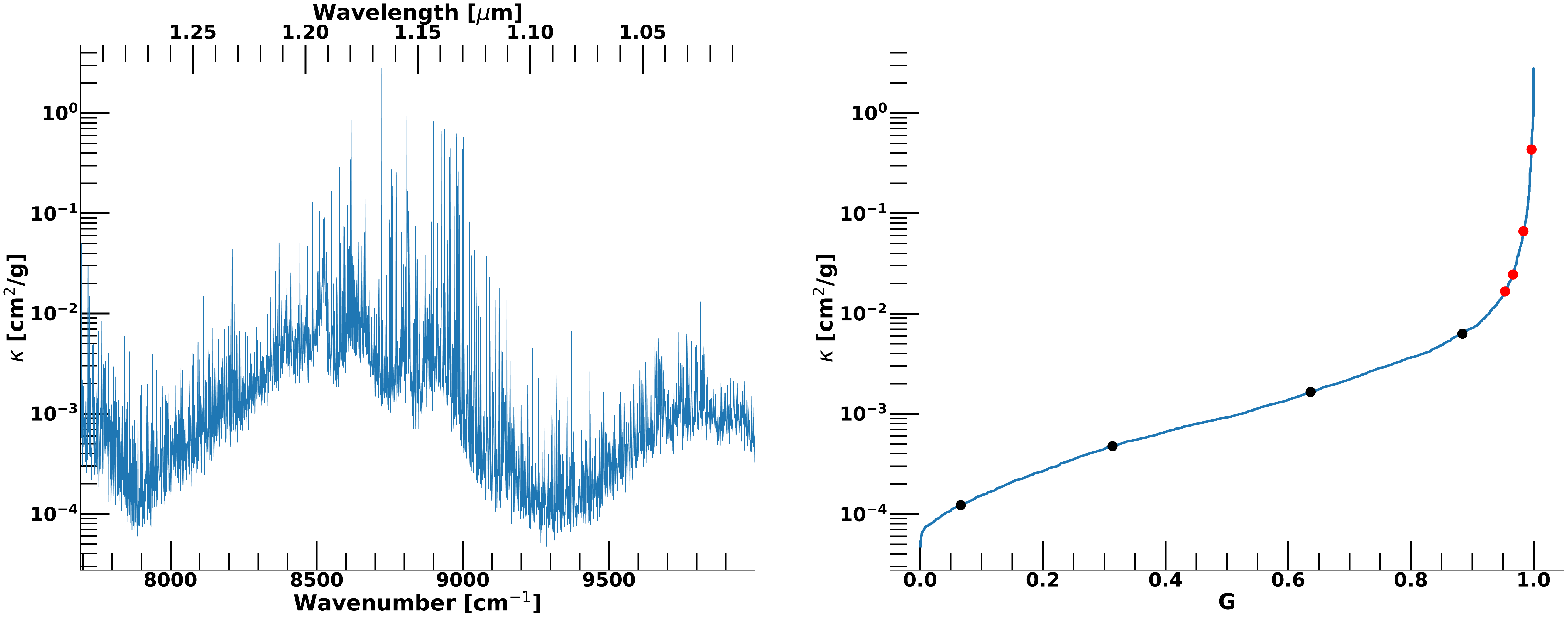}
  \caption{{\bf Left panel} shows the high resolution opacity $\kappa$ between 1 and 1.3 $\mu$m of a solar mixture of atmospheric gases at a temperature of 550 K and a pressure of 80 mbar. {\bf Right panel} shows the k-distribution computed for the opacity shown in the left panel with the blue solid line. The black and red points are the Gauss points (from Table \ref{table:tab2}) overplotted on the k-distribution for this opacity window. These Gauss points are applicable for the double Gauss point method. The first set of 4 Gauss points are shown in black while the second set is shown with red points. }
\label{fig:figkcoeff}
\end{figure*}

In the correlated-k approach, the relevant wavelength range for the radiative transfer is first divided to 196 wavelength bins. These carefully chosen wavelength bins are shown in Figure \ref{fig:figwbin} and chosen to reflect the approximate spectral energy distributions of planetary atmospheres. The molecular opacity $\kappa$ in any one of these wavelength bins has numerous individual molecular lines, as  shown in Figure \ref{fig:figkcoeff} left panel. The cumulative distribution function (CDF) of the opacity can be defined as $G(\kappa_0) = N(\kappa \le \kappa_0)$ where $N$ denotes the total number of instances when the condition $\kappa \le \kappa_0$ is satisfied within the wavelength bin. The CDF $G(\kappa)$ of the opacity within each wavelength bin is computed. This function, $G(\kappa)$, is then inverted to obtain the ``k-distribution''. An example  k-distribution for the the opacity window shown in the left panel in Figure \ref{fig:figkcoeff}, is shown in Figure \ref{fig:figkcoeff} right panel. After computing a k-distribution for each of the 196 wavelength bins, the 8 Gauss points ($g_i$) shown in Table \ref{table:tab2} are used to represent each distribution. Generally, the slope of k-distributions are shallow and slowly changing at lower Gauss point values ($G$) between 0 and 0.9. Then, the slope steepens  rapidly between $G$= 0.9--1, as can be seen in Figure \ref{fig:figkcoeff}. To capture the complex shape of the k-distribution with just 8 Gauss points, a double-Gauss method is used for the integration. The first set of four Gauss points are  shown in Figure \ref{fig:figkcoeff} right panel with black points. These four Gauss points sparsely sample most of the k-distribution between $G=$0--0.95 whereas the last four Gauss points, shown with red points in Figure \ref{fig:figkcoeff}, sample the small, but rapidly changing, part of the k-distribution between $G=$0.95--1. The first and second set Gauss points and weights for this double-Gauss point method can be generated from sample points of the generally used Gauss-Legendre quadrature \citep{abramowitz} using,
\begin{equation}
\begin{aligned}
    g_{i,1} = & f(g_i^{'} + 1)/2 \\
    w_{i,1} = & fw_i^{'}/2 \\
    g_{i,2} = & f+ (1-f)(g_i^{'} + 1)/2 \\
    w_{i,2} = & (1-f)w_i^{'}/2 \\
\end{aligned}
\end{equation}

where  $g_{i,1}$ and $w_{i,1}$ represent the first set of our Gauss points (shown with black points in Figure \ref{fig:figkcoeff}) and $g_{i,2}$ and $w_{i,2}$ represents the second set (shown with red points in Figure \ref{fig:figkcoeff}), $g_i^{'}$ and $w_i^{'}$ are the sampling points for Gauss-Legendre quadrature of some order (4 in our case) defined within the interval [-1,1], and $f$ is the adjustable parameter which sets the division in the values of $G$ which will be sampled by the first and second set of Gauss points. For example, if $f=0.8$ then $g_{i,1}$ will sample values between 0 and 0.8 while the rest will be sampled by $g_{i,2}$. For our purpose, this $f$ is set at 0.95 which is a reasonable choice based on the typical shape of k-distributions as shown in Figure \ref{fig:figkcoeff}.

This double-Gauss method for calculating correlated k-coefficients has a large impact on the time and computational efficiency of the code as it reduces the number of Gauss points required for radiative transfer calculations but still maintains sufficient accuracy required for the model. It has been shown that the radiative transfer with double-Gauss method with 8 Gauss points is as accurate as a normal set of 20 Gauss points \citep[Michael Line, priv. comm. as detailed in][]{marley21}. The choice of the wavelength-bins for the correlated-k opacities are also crucial for the model and are described next.

The 196 wavelength bins used span 0.2 $\mu$m -- 227 $\mu$m to capture the general spectral energy distribution of planetary atmospheres. However, these wavelength bins are not all of equal wavelength width. These wavelength bins are shown in Figure \ref{fig:figwbin} where three blackbody curves corresponding to temperatures of 300 K, 1000 K and 2000 K are also shown for comparison. The bins are narrow between 0.5-10 $\mu$m to capture the large number of molecular rovibrational bands in this range. At the tail ends of the blackbody distributions, larger bins help boost computational speed and maintain sufficient precision in radiative transfer calculations required for the application cases of this model.

\begin{figure*}
  \centering
  \includegraphics[width=1\textwidth]{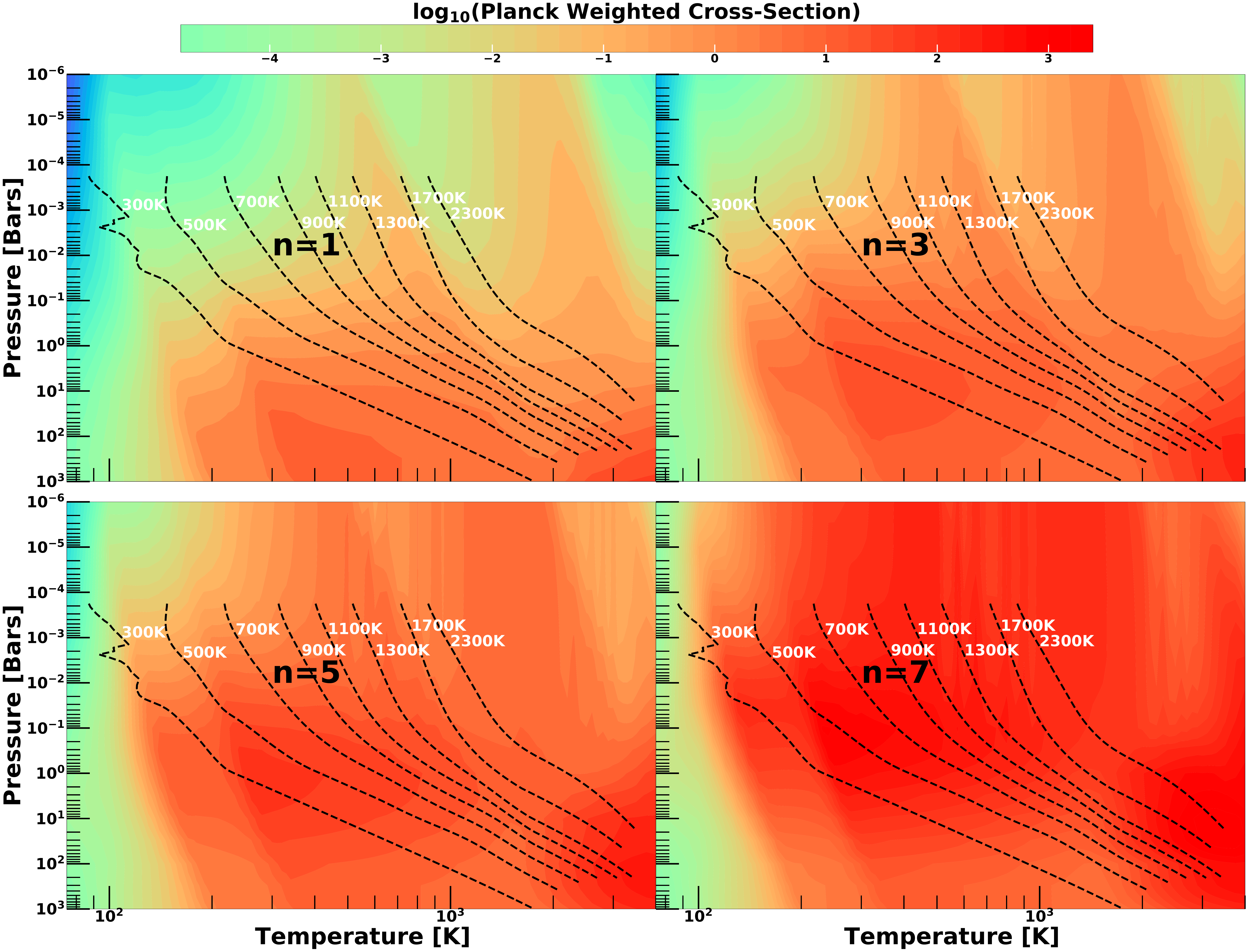}
  \caption{{\bf Top left panel} shows the heat map of planck weighted molecular opacity in the pressure-temperature k-coefficient grid used in the model with equilibrium chemistry described in the paper. Dashed black lines marked on the cross-section map depict different converged $T(P)$ profiles of brown dwarfs from \citet{marley21} with {\teff} between 300 K and 2300 K and log(g)=5. This grid is used for interpolation of opacities during iteration of the atmospheric state in the code. The same for the 3rd, 5th, and 7th Gauss points are also shown in the {\bf top right}, {\bf bottom left} and {\bf bottom right} panel. }
\label{fig:figpremixopa}
\end{figure*}

As our model uses a pre-computed chemistry grid during iterations on the $T(P)$ profile, the k-coefficients are also pre-computed on the same pressure--temperature--metallicity--C/O ratio grid. For intermediately $T-P$ values, the molecular opacity is interpolated using the same formalism as has been described for the gas abundances in \S\ref{sec:chem}. Figure \ref{fig:figpremixopa} shows heat maps of pre-computed Planck function and abundance weighted molecular cross-sections in the same pressure-temperature grid as Figure \ref{fig:figabun}. For each $P-T$ point in the grid, the abundance weighted molecular cross-sections are integrated over all wavelengths using the Planck function (corresponding to the temperature $T$) as the integrating kernel (see Equation 2 in \citet{freedman14}). Each panel corresponds to cross-sections at different Gauss points which we use in our models. Planck mean cross sections at only the first four Gauss points are shown here in the four panels. Dashed black lines marked on the cross-section maps depict different converged $T(P)$ profiles of brown dwarfs from \citep{marley21} with {\teff} between 300 K and 2300 K and log(g)=5. As the order of the Gauss points increases in Figure \ref{fig:figpremixopa}, the cross-sections become higher as higher order Gauss points trace higher opacity parts of the k-distribution. Achieving atmospheric convergence in regions of the $P-T$ space where the cross-section change rapidly with small changes in temperature or pressure can be difficult. Such an ``opacity cliff" can be seen at $T$ values between$\sim$ 900-1700 K in Figure \ref{fig:figpremixopa}. This cliff appears in the Planck mean cross-sections due to the overlap of the peak of the Planck function at 900-1700 K with large {\water} opacity bands between 1.5-3 $\mu$m. These ``opacity cliffs" were also seen in \citet{freedman2008opacities,freedman14} with Rosseland-mean opacities. We discuss the effect of these cliffs on model convergence in more detail in \S\ref{sec:convergence_rec}. 

In addition to molecular opacities, collision induced absorption (CIA) of H$_2$-H$_2$, H$_2$-H, H$_2$-He, H$_2$-N$_2$, H$_2$-CH$_4$, and continuum opacities such as H-bf, and H-ff are also accounted for. These are included separately from our correlated-k table. The CIA opacities are pre-calculated between temperatures of 75 K and 7000 K with 1000 wavelength bins and then interpolated to the correct $P-T$ combination with a spline. Note that even though the CIA temperatures are computed up to 7000~K, the temperature valid range of our model is still limited by the 1460 opacity grid.

We note that the pre-computed opacity grid used with this model is known to be incomplete with regards to atomic and ionic opacities, which are particularly important at the high temperature and low pressure parts of the atmosphere \citep[e.g.][]{hoeijmakers2018atomic, hoeijmakers2020hot}. %This is generally not an issue for the high temperature and high pressure atmosphere because these parts are generally convective. 
This limits, to some extent, the current code's ability to treat the ``ultra-hot Jupiters" ($T_\mathrm{eq}\gtrapprox$2200~K). In a future update we will add these opacities to the correlated-K tables.

\subsubsection{Disequilibrium Chemistry}\label{sec:deq_chem}

A significant addition to this model is the capability to treat vertical mixing induced disequilibrium chemistry self-consistently within the radiative-convective equilibrium framework. This is an optional part of the model, as shown in Figure \ref{fig:figschematic}. In 1D models, vertical mixing is often parametrized as a diffusive process which is described by the eddy diffusion coefficient, {\kzz} \citep{allen81}. The mixing timescales of all relevant gases in each atmospheric layer is given by,
\begin{equation}\label{eq:tmix}
    t_{\rm mix} = \dfrac{H^2}{K_{\rm zz}}
\end{equation}
where $H$ is the local scale height of that atmospheric layer. All thermochemical reactions within the atmosphere like CO$\leftrightarrow$CH$_4$ and NH$_3${$\leftrightarrow$}N$_2$ also proceed with a characteristic timescale -- {\tchem}. \citet{Zahnle14} parametrized the {\tchem} of several such gases using 1D chemical kinetics models. The parametrized {\tchem} from \citet{Zahnle14} and \citet{visscher06} are given by,

\begin{align*}
        t_{\rm CO,CH_4,H_2O} &= \dfrac{3\times10^{-6}}{P_{\rm bar}}exp\left(\dfrac{42000 K}{T}\right)\\ 
       t_{\rm NH_3,N_2} &= \dfrac{10^{-7}}{P_{\rm bar}}exp\left(\dfrac{52000 K}{T}\right) \\
       t_{\rm CO_2} &= \dfrac{10^{-10}}{\sqrt{P_{\rm bar}}}exp\left(\dfrac{38000 K}{T}\right)\\
       t_{\rm HCN} &= \dfrac{1.5\times10^{-4}}{P_{\rm bar}m^{0.7}}exp\left(\dfrac{36000 K}{T}\right)\\
       t_{\rm PH_3} &= \dfrac{1.9\times10^{12}}{\rm [OH]}exp\left(\dfrac{6013.6 K}{T}\right) 
\end{align*} \label{eqn:tchem}
where $P_{\rm bar}$ is the atmospheric pressure in bars, $T$ is the temperature and $m$ is the atmospheric metallicity relative to solar metallicity. {\tchem} is generally short at high pressure-high temperature regions of the atmosphere. As the atmosphere gets colder at lower pressures the {\tchem} rises exponentially and becomes  large. The volume mixing ratios of {\co}, {\meth}, {\amon}, {\cotwo}, {\water} and HCN are expected to follow equilibrium chemistry at pressures where {\tmix} $>>$ {\tchem} which happens in the deeper atmosphere. But as the atmosphere becomes colder with lowering pressure, {\tchem} may exceed {\tmix}. The pressure at which this occurs is called the ``quench pressure". At pressures less than the ``quench pressure", gases are expected to depart from chemical equilibrium and their mixing ratios  become constant.

In this model, we include these disequilibrium chemistry effects for the 8 gas species shown in Eqn. \ref{eqn:tchem}: {\water}, {\meth}, {\co}, {\cotwo}, {\amon}, N$_2$, HCN, and PH$_3$. The quench pressure is determined using the  parametrized {\tchem}  in Eqn. \ref{eqn:tchem}. The abundance then follows the equilibrium chemistry at pressures more than the quench level, and is held constant at pressures less than the quench level.

The uncertain parameter in the disequilibrium modeling framework is the eddy diffusion parameter, {\kzz} \citep{Philips20,fortney20,karilidi21}. Therefore, this model is  flexible in assumptions regarding {\kzz}. Currently, there are two user-options for defining \kzz\ within a model run: 1) a fully user-defined {\kzz} value which is either constant or variable throughout the height of the atmosphere, but which does not change during the iterative climate solution  \citep[as was done in][]{karilidi21,Philips20}, and 2) a model-predicted  {\kzz}, which is  calculated from the $T(P)$ profiles in the convective zones, using mixing length theory \citep{gierasch85energy}, and in the radiative zones, using parametrizations \citep[e.g.][]{moses21}. In the second case, along with the $T(P)$ profile, the \kzz\ will also change simultaneously with the iterations in the model. The {\kzz} in the convective zone can be calculated with mixing length theory and is given by \citep{gierasch85energy},

\begin{equation}\label{eq:kzz_con}
 K_{zz}= \dfrac{H}{3}{\left(\dfrac{L}{H}\right)}^{{4}/{3}}{\left(\dfrac{RF}{\mu{\rho_a}c_p}\right)}^{{1}/{3}}
\end{equation}

where $H$ is the local scale height of the atmosphere, $L$ is the turbulent mixing length, $R$ is the universal gas constant, $\mu$ is the mean molecular weight of the atmosphere, $\rho_a$ is the atmospheric density, $c_p$ is the atmospheric specific heat at constant pressure and $F$ is the convective heat flux. The convective heat flux can be calculated by the difference between the net thermal radiative flux and $\sigma{T_{\rm eff}}^4$ within the convective zones of the atmosphere. Therefore, $\sigma{T_{\rm eff}}^4$ is the maximum allowed value of $F$ in this framework if the energy transport within the convective atmosphere is assumed to be completely convective.

While various parametrizations of {\kzz} in the radiative zones of substellar atmospheres have been discussed in the literature \citep{wang15,zhang18,parmentier13,tan22}, we have included the parametrization from \citet{moses21} as a starting point. Further parametrizations of {\kzz} can easily be swapped in, in the future. The \citet{moses21} radiative zone {\kzz} is given by,

\begin{equation}\label{eq:kzz_rad}
    K_{zz} = \dfrac{5\times10^{8}}{\sqrt{P_{\mathrm{bar}}}}\left(\dfrac{H}{{620}  {\rm km}}\right)\left(\dfrac{T_{\rm eff}}{1450 {\rm K}}\right)^{4}
\end{equation}
where $P_{\mathrm{bar}}$ is the pressure of the radiative level in bars and $H$ is the atmospheric scale height. Both Equation \ref{eq:kzz_con} and \ref{eq:kzz_rad} can be found in the \href{https://github.com/natashabatalha/picaso/blob/caf63752563215e76ec713f65182f7efc367f3fc/picaso/fluxes.py#L3455}{\texttt{get\char`_kzz}} function. As the $T(P)$ profile iterates towards the converged solution, the quench levels of various gases  change as well. This means that the abundances of quenched species will depart from chemical equilibrium, and the pre-computed k-coefficient tables described in \S\ref{sec:opa} are no longer valid. The methodology to remix the k-coefficients with updated abundances and recompute resulting optical depth calculation during the convergence process, is referred to ``on--the--fly" mixing, and is described in the following \S\ref{sec:onthefly}.

\subsubsection{Mixing Opacities ``on--the--fly"}\label{sec:onthefly}
With disequilibrium chemistry, the atmospheric chemistry depends on the quench pressures, which again depend on the $T(P)$ profile of the atmosphere. Therefore, because the chemistry of the atmosphere cannot be pre-determined, the atmospheric opacities also need to be calculated ``on--the--fly".

We mix the correlated-k opacities of individual gases using the methodology of \citet{amundsen17} called the resort-rebin technique. Currently, in this model we focus on the quenching of {\co}, {\meth}, {\water}, {\amon}, {\cotwo}, N$_2$, HCN and PH$_3$. However, the major opacity sources among these gases are mainly {\co}, {\meth}, {\water}, and \amon. Meaning, the  contribution of N$_2$, HCN and PH$_3$ on the total gas opacity is negligible for the relatively small departures from the chemical equillibrium with $\log$(M/H)$\le$2.5 explored in this analysis. %For example, the quenched abundance of, PH$_3$ is not expected to change by such a quantity that would make PH$_3$ the dominant opacity contributor at any wavelength within our grid.    
Therefore, we mix the correlated-k opacities of {\co}, {\meth}, {\water}, and {\amon} with the correlated-k opacities of all the other sets of background gases which follow equilibrium chemistry as the volume mixing ratio of these gases evolve due to quenching with the $T(P)$ profile of the atmosphere. Should the motivation to include more gases in the ``on--the--fly'' methodology present themselves in future observations, they can be included in a future code release.

One of the drawbacks of  ``on--the--fly mixing" outlined in \citet{amundsen17} is the dependence of the accuracy of the technique on the spectral resolution of the correlated-k opacities. This effect has been well explored in \citet{karilidi21}. Specifically, it was found that the 196 wavelength bins,  traditionally used in the chemical equilibrium  version of \texttt{EGP} and this new python version, is not sufficiently high to use with the resort-rebin technique. \citet{karilidi21} found that 661 wavelength bins in  the correlated-k opacities are required to counteract the inaccuracies in the resort-rebin technique. Therefore, when disequilibrium calculations are requested by the user, \texttt{PICASO} automatically switches to  661 wavelength bins. 

With the opacities and mixing routines described, the final module, shown in Figure \ref{fig:figschematic} is the computation of the convective zones, which we outline in the following \S\ref{sec:convec}.

\subsubsection{Convective Zones}\label{sec:convec}

\begin{figure}
  \centering
  \includegraphics[width=0.48\textwidth]{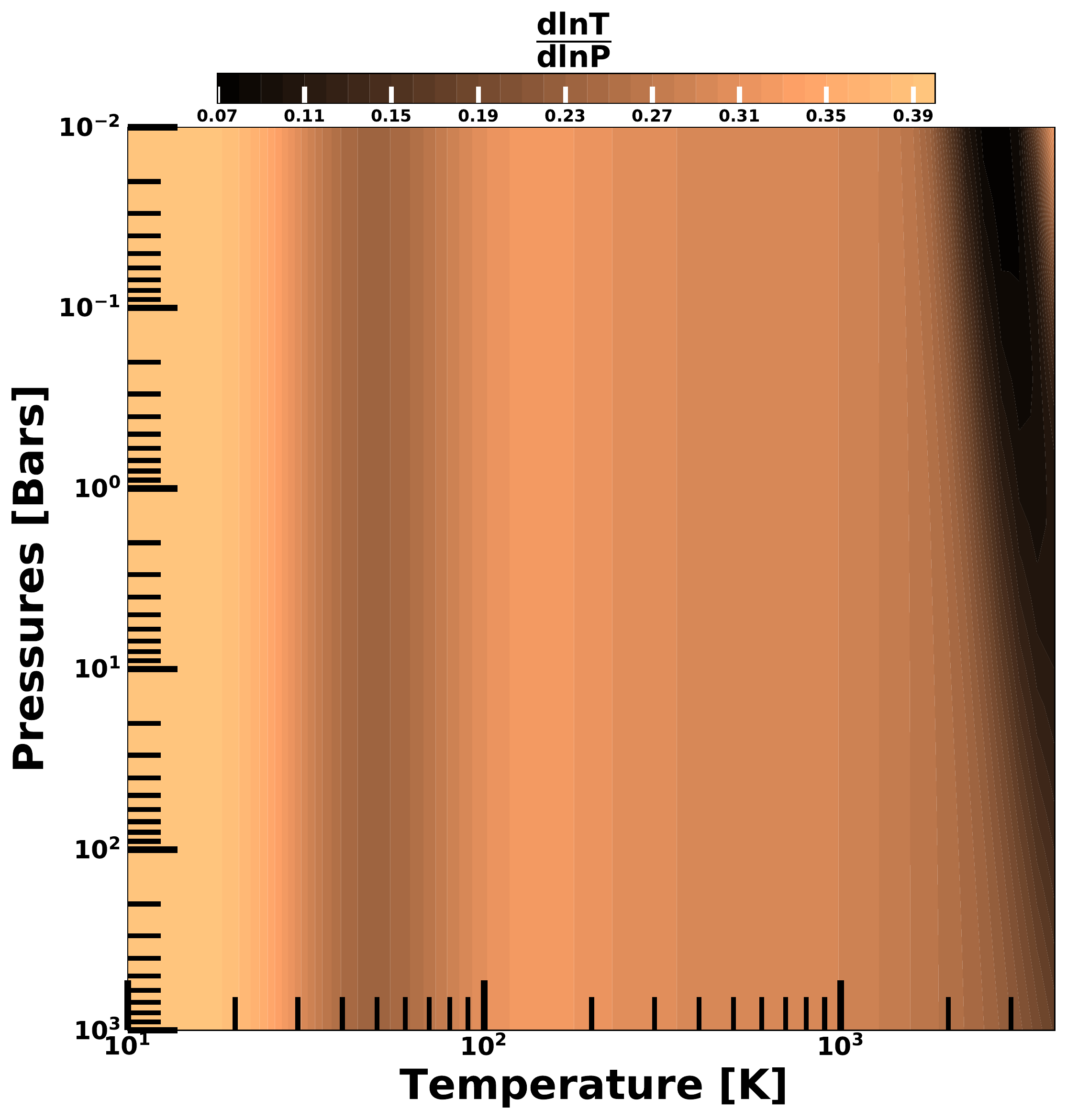}
  \caption{A heat map of the adiabatic lapse rate as a function of pressure and temperature is shown here. This grid is used to determine the lapse rate in atmospheric layers unstable against convection. }
\label{fig:figconvec}
\end{figure}

As the $T(P)$ profile iterates towards radiative equilibrium, parts of the atmosphere will  become unstable against convection, and energy transport will be expected to occur via convection instead of radiation. These parts of the atmosphere are forced to follow the local adiabat. Figure \ref{fig:figconvec} shows a heat map of the adiabatic lapse-rate, ${\rm dlnT}/{\rm dlnP}$, used in \texttt{PICASO}. This grid is pre-computed assuming a solar H--He mixture with He mass fraction of Y=0.28. This includes H$_2\leftrightarrow$ 2H dissociation and a detailed accounting of the molecular vibrational and rotational levels\footnote{Raw grid can be found in GitHub file \href{https://github.com/natashabatalha/picaso/blob/climate/reference/climate_INPUTS/specific_heat_p_adiabat_grad.json}{specific\_heat\_p\_adiabat\_grad.json}}. The calculations for this grid was done by Didier Saumon as described in \citet{marley21}. The pre-computed grid has 53 temperature points between 10 K and 3981 K and 26 pressure points between 10$^{-2}$ and 10$^3$ bars. The lapse-rate of parts of the profile which become unstable against convection are interpolated from this grid. For every atmospheric layer, the lapse-rate, ${\rm dlnT}/{\rm dlnP}$,  is first calculated from the $T(P)$ profile. The grid of adiabatic lapse-rates shown in Figure \ref{fig:figconvec} is then used to interpolate the local adiabatic lapse rate for each atmospheric layer. This interpolation is done using 2D interpolation similar to the technique described in \S\ref{sec:chem}. 

If the ratio between the lapse-rate obtained from the $T(P)$ profile and the interpolated, lapse rate, $\nabla$ is greater than $\sim$1 (numerically set to 0.98) in any of the layers, then these layers are considered convective and are forced to follow the interpolated local adiabatic lapse rate according to the following equation,

\begin{equation}
    T^{i+1} = \exp\left({T^{i} +{\nabla}ln\left(\dfrac{P^{i+1}}{P^{i}}\right)}\right)
\end{equation}

where $T^{i}$ and $P^{i}$  are the temperature and pressure of the i'th convective layer and $T^{i+1}$ and $P^{i+1}$ are the temperature and pressure of the $i$+1'th convective layer starting from below the radiative-convective boundary layer. We should note that this convective adjustment is done for one layer at a time followed by a iteration of the entire $T(P)$ as this adjustment for each layer, in principle, would lead to temperature changes in all the other layers. This approach used in this model is different from forcing all the layers which are unstable against convection to be convective at once.  The \href{https://github.com/natashabatalha/picaso/blob/caf63752563215e76ec713f65182f7efc367f3fc/picaso/climate.py#L603}{\texttt{t\char`_start}} module implements this convective adjustment of the $T(P)$ profile. This methodology is used to develop and grow convective zones in the atmosphere with the \href{https://github.com/natashabatalha/picaso/blob/caf63752563215e76ec713f65182f7efc367f3fc/picaso/justdoit.py#L3589}{\texttt{find\char`_strat}} module  during the iterations of the model. Note that this module is designed to always form or grow convective zones and not shrink or remove them. We discuss the important implications of this for the user in \S\ref{sec:recommendations}.

\subsection{Iteration Scheme And Model Convergence}\label{sec:model_convergence}
Each required and optional physical component of the model shown in Figure \ref{fig:figschematic} has been described separately in \S\ref{sec:physics}. But as our model is iterative in nature, a description of the iterative scheme of the model along with its convergence criteria is a necessary component. We provide a brief outline of the model iterative scheme for both the chemical equilibrium and disequilibrium runs separately below.

\subsubsection{Iterative Scheme for Chemical Equilibrium}\label{sec:chem_eq_convergence}

This model aims to achieve radiative-convective equilibrium using the Newton-Rhapson iterative scheme. As a first step, the chemistry and the correlated-k opacities of the atmosphere for the first given guess $T(P)$ profile are calculated as detailed in \S\ref{sec:chem} and \ref{sec:opa}. Then, the methodology described in \S\ref{sec:RT} is used to compute the net radiative fluxes in all atmospheric layers. The temperature of each layer is then perturbed by a small arbitrary $dT$, historically set at 0.01\% of the current layer temperature, while keeping the other layer temperatures fixed. The radiative fluxes in all layers are then recomputed after this perturbation. These two sets of radiative fluxes are used to compute the Jacobian A$_{ij}$ which quantifies the change in radiative flux in layer $i$ due to perturbed temperature in layer $j$. The Jacobian is given by \citep[e.g.][]{hubeny17},
\begin{equation}
    A_{ij} = \dfrac{F^{j}_{i,p}-F_{i}}{dT_{j}}
\end{equation}
where $F_{i}$ is the net radiative flux in layer $i$ before the perturbation and  $F^{j}_{i,p}$ is the perturbed net radiative flux in layer $i$ due to a change in temperature $dT_{j}$ in layer $j$. The ultimate temperature correction, $\delta{T}$, needed to ensure radiative equilibrium (in the radiative zones) is then solved using the equation,
\begin{equation}\label{eq:correction}
    A\delta{T} = \sigma{T_{\rm eff}}^4 - F(T(P))
\end{equation}
where $A$ is the Jacobian and $F(T(P))$ are the net radiative fluxes in each layer with the current atmospheric state. % and $\delta$T are the needed corrections to the T(P) profile. 
This process continues iteratively until a tolerance of maximum allowed radiative flux difference from  $\sigma$\teff$^4$ is reached in all the radiative layers of the atmosphere. This convergence criteria can be expressed as,

\begin{equation}\label{eq:tolerance}
    \left|\dfrac{F^n_{\rm net} - {\sigma}T_{\rm eff}^4}{{\sigma}T_{\rm eff}^4}\right| \le \epsilon
\end{equation}
where $\epsilon$ denotes the tolerance parameter, set at $5\times10^{-3}$ in our model. This equation is a numerical version of Equation \ref{eq:RCEfin}. The model is also considered to be converged if the maximum fractional temperature correction among all the radiative layers $|\delta{T/T}|$, calculated from Equation \ref{eq:correction}, during any iteration is smaller than $\epsilon$. It is important to note that the model satisfies the convergence criteria in Equation \ref{eq:tolerance} in the mid-point of each atmospheric layer while the temperature at the edges of each atmospheric layer (levels) is iterated. For example, if the optical depths at the bottom and top (edges) of the $i$'th atmospheric layer are $\tau_i$ and $\tau_{i-1}$, then the model aims to satisfy Equation \ref{eq:tolerance} in this layer for an optical depth of $(\tau_i+\tau_{i-1})/2$. But, the iteration of the temperature profile is done to the temperature at the bottom and top (or edges) of this atmospheric layer. This is important for the stability of the iterative process.

As the iteration progresses, the stability of each atmospheric layer against convective mixing is also checked using the technique described in \S\ref{sec:convec}. If layers unstable against convection are found, the temperature of these convectively unstable layers are forced to a H$_2$-He gas mixture adiabat from the grid shown in Figure \ref{fig:figconvec}. Once the convergence criteria of Equation \ref{eq:tolerance} is met, the model run stops and it produces the outputs outlined by the green boxes in Figure \ref{fig:figschematic}. Outputs such as the converged $T(P)$ profile, chemical abundances and outgoing radiative fluxes in the 196 wavelength grid are always returned to the user after model converges. These outputs can further be optionally used for calculating various observables such as  transmission, emission or reflected light spectra with \texttt{PICASO}.

\begin{figure*}
  \centering
  \includegraphics[width=1\textwidth]{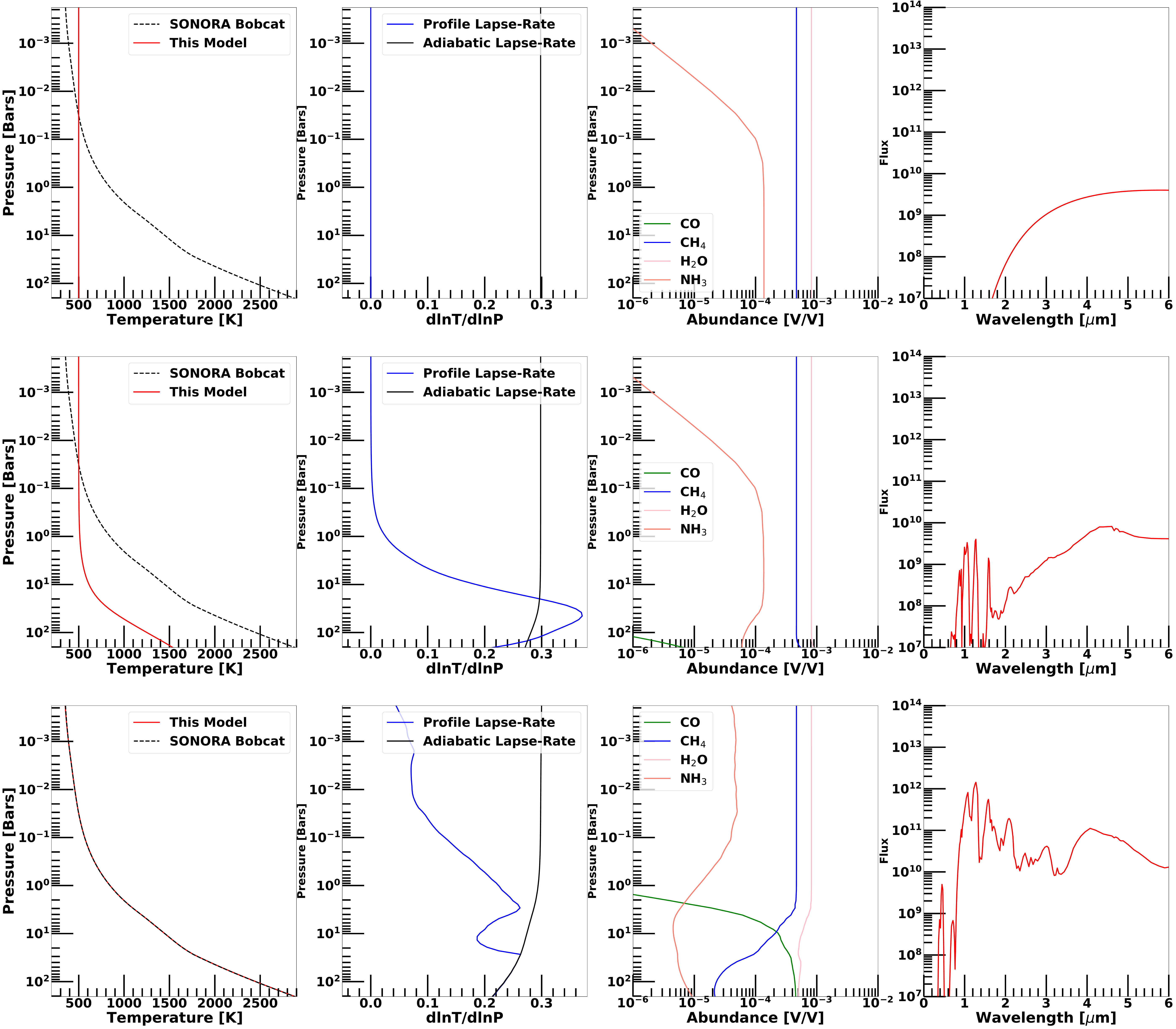}
  \caption{Three steps in the iteration towards a converged solution for a brown dwarf with {\teff} of 1000 K and log(g) of 5 is shown here in each row. The first  column shows the iteration of the $T(P)$ profile in red with the \texttt{SONORA BOBCAT} model shown in black dashed. The second  columns show the lapse-rate from the $T(P)$ profiles in blue and the local adiabatic lapse rate in black. The third column shows the iterations in the volume mixing ratios of various gases. The fourth column shows the thermal spectrum emergent from the atmosphere of the object in each iteration. An animated gif of this figure can be found \href{https://drive.google.com/file/d/1pEb8Ih9GVliCMlPuL_ZgUGXgRHAyTt1r/view?usp=sharing}{here}  }
\label{fig:figiter_bd}
\end{figure*}

\begin{figure*}
  \centering
  \includegraphics[width=1\textwidth]{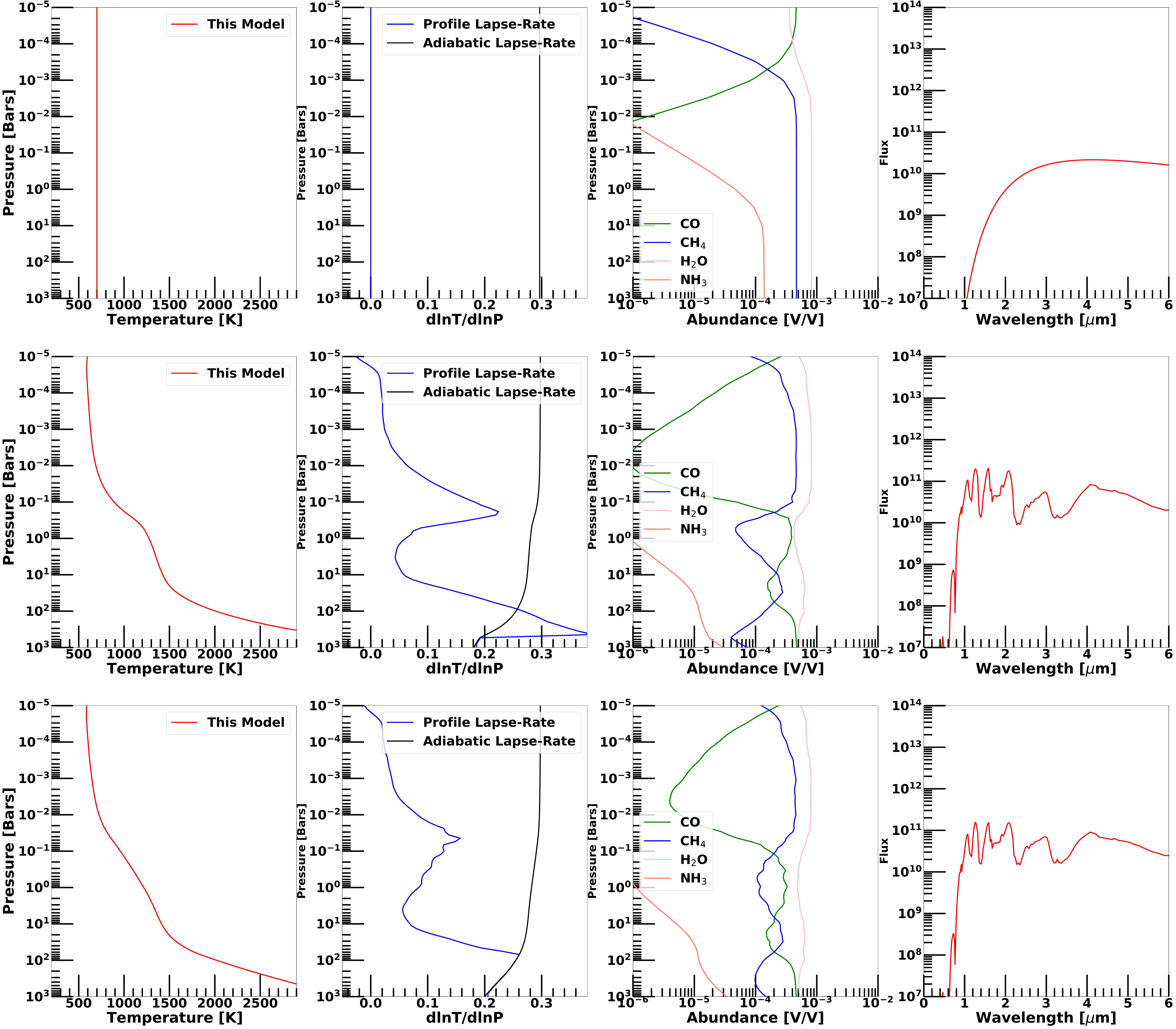}
  \caption{Three steps in the iteration towards a converged solution for an exoplanet around a sun like star with {\tint} of 300 K and log(g) of 3.4 is shown here in each row. The left-most column shows the iteration of the $T(P)$ profile in red with the \texttt{SONORA BOBCAT} model shown in black. The second left-most columns show the lapse-rate from the $T(P)$ profiles in blue and the local adiabatic lapse rate in black. The third column shows the iterations in the volume mixing ratios of various gases. The right-most column shows the thermal spectrum emergent from the atmosphere of the object in each iteration. An animated gif of this figure can be found \href{https://drive.google.com/file/d/1bwm2r48SN75--aRAn2uHVyCvbcLcLjxp/view?usp=sharing}{here}  }
\label{fig:figiter_exo}
\end{figure*}

For a deeper understanding of the iteration scheme of this model, three atmospheric iteration steps for a brown dwarf with {\teff} of 1000 K and log(g)=5 are shown in Figure \ref{fig:figiter_bd} and iterations for an irradiated planet at a distance of 0.1 AU from a sun like star with {\tint} of 300 K and log(g)=3.4 are shown in Figure \ref{fig:figiter_exo}. In Figure \ref{fig:figiter_bd} and \ref{fig:figiter_exo}, each row corresponds to the atmospheric state in a certain iteration of the model. The first column shows the $T(P)$ profile of the atmosphere in each iterative step of this model with the red solid line whereas \texttt{SONORA} $T(P)$ profile for this case is shown as a black dashed line. The second column shows the lapse-rate of the $T(P)$ profile as a function of pressure in each iteration with the solid blue line and the adiabatic lapse-rate is shown with a black solid line. The third column shows the volume mixing ratio profiles of four gases -- {\water}, {\meth}, {\amon}, and {\co} as a function of pressure and the fourth column shows the emergent thermal spectrum from the top of the atmosphere. In Figure \ref{fig:figiter_bd}, the first row shows a simple initial isothermal guess profile of 500 K and the last row shows the final converged solution. For our initial guess, the bottom four layers of the atmosphere have been assumed to be convective. The initial thermal emission spectrum resembles a blackbody because the atmosphere is isothermal. As the model iterates, the isothermal $T(P)$ profile is perturbed to reach an atmospheric state such that Equation \ref{eq:tolerance} is satisfied for all radiative layers of the atmosphere. In the iteration shown in the middle row,  the bottom of the atmosphere becomes unstable against convection and therefore in the final solution the lapse-rate in that part of the atmosphere is forced to follow the local adiabatic lapse-rate. As the model iterates towards the converged solution, the $T(P)$ profile becomes more complex and leads to a redistribution of flux in each wavelength bin. This leads to molecular absorption lines becoming more evident in the thermal spectrum with each iteration. The final converged solution matches  well with the \texttt{SONORA BOBCAT} grid of models (shown as black dashed line). For the irradiated planet convergence shown in Figure \ref{fig:figiter_exo}, the initial guess is an isothermal profile with $T= 700$ K and the iterations continue until Equation \ref{eq:tolerance} is satisfied.

\begin{figure*}
  \centering
  \includegraphics[width=1\textwidth]{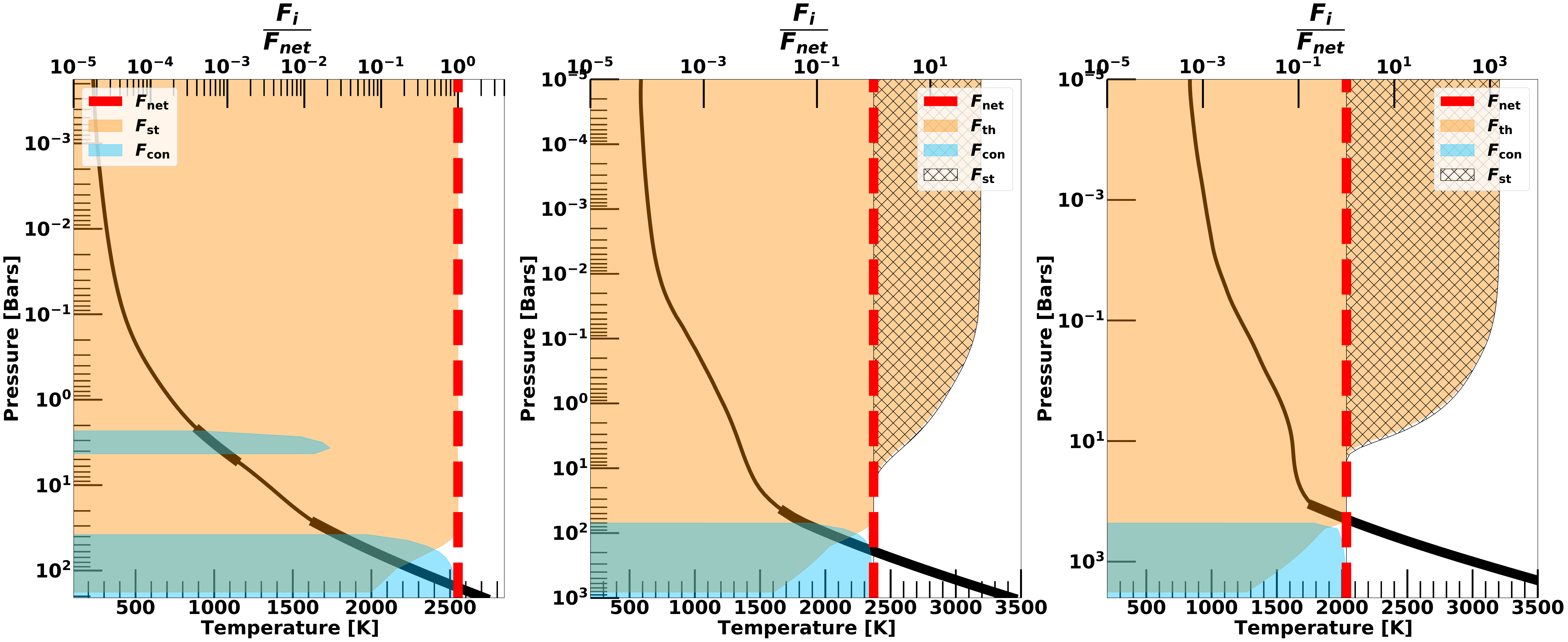}
  \caption{Radiative and convective flux contributions compared to total net fluxes are shown for three  $T(P)$ profiles in order to visualize the concept of ``convergence''. From left to right a representative brown dwarf, warm Jupiter, and hot Jupiter profile is shown. In each figure, the shaded blue region depicts the convective flux at each layer. The orange shaded region depicts the net radiative flux in each layer. The red dashed line shows the target flux for the brown dwarf in the left panel. For the brown dwarf (non-irradiated) this target flux is $\sigma$\teff$^{4}$. The red dashed line in the middle and the right panels shows the internal heat flux of the planets which is $\sigma$\tint$^{4}$. Please note that the scale of the upper and lower X-axis are different in all the three panels.}
\label{fig:fig_bd_exo1_exo2}
\end{figure*}

A second helpful method of visualizing a converged run is to look at the contribution of radiative and convective fluxes as compared to the target flux in brown dwarfs given by $\sigma$\teff$^{4}$ or the internal heat flux for planets given by  $\sigma$\tint$^{4}$. Figure \ref{fig:fig_bd_exo1_exo2} shows these layer--by--layer radiative and convective fluxes for converged models of a representative brown dwarf (left panel), warm Jupiter (middle panel), and hot Jupiter (right panel). The net radiative flux in each layer is shown with the orange shaded region. The shaded blue region depicts the convective flux at each layer. The red dashed line shows the target flux for the brown dwarf -- $\sigma$\teff$^{4}$. This converged brown dwarf model has two convective and radiative zones as can be seen from its $T(P)$ profile shown in black. As a result, the convective flux (blue) peaks at the location of these convective zones. The net radiative flux decays rapidly at the deeper convective zone and convection carries the majority of the energy in these deep convective layers. The proof that this is a converged model and Equation \ref{eq:tolerance} is satisfied lies in the fact that the sum of the net radiative and convective fluxes is equal to the target flux at all atmospheric layers. For the planet case where there is an additional  energy flux from the host star, we show the additional energy with black hatched shading (2nd and 3rd column of Figure \ref{fig:fig_bd_exo1_exo2}). This incident flux is downward, compared to the thermal radiative and convective fluxes, which is upward. Therefore, the upper atmosphere needs larger quantities of upward thermal radiative fluxes to balance this downward stellar flux. Ultimately, the goal is to maintain a summed flux of $\sigma$\tint${^4}$ (red dashed line) at all the atmospheric layers of the planet. Therefore, the upper atmosphere of irradiated planets, with a given \tint\ , is pushed toward hotter temperatures when compared to a  brown dwarf with a comparable {\teff}.

\subsubsection{Iteration Scheme for Chemical Disequilibrium}\label{sec:EGP_tk}

\begin{figure*}
  \centering
  \includegraphics[width=1\textwidth]{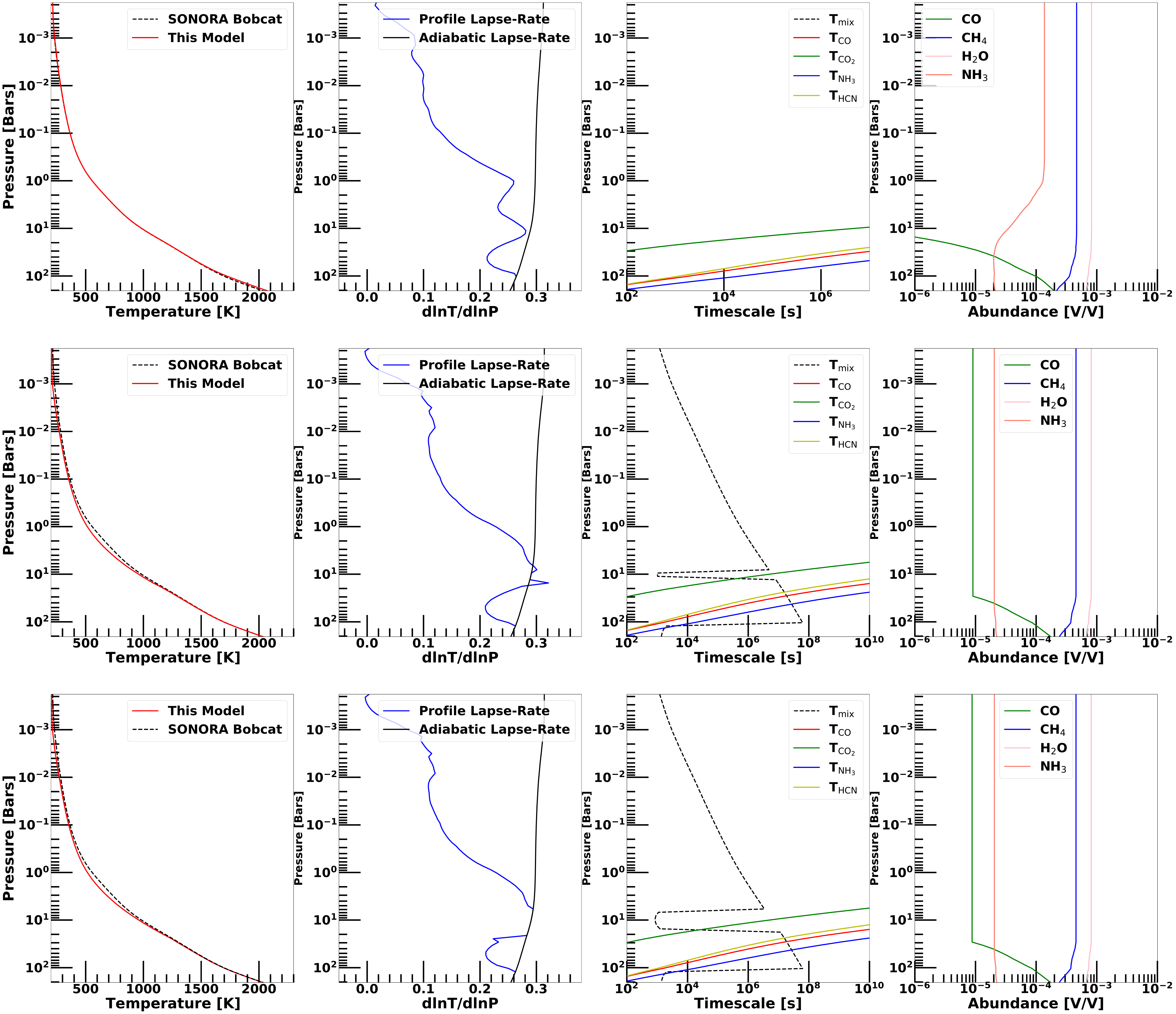}
  \caption{Three steps in the iteration towards a converged solution with disequilibrium chemistry for a brown dwarf with {\teff} of 700 K and log(g) of 5.25 is shown here in each row. The first  column shows the iteration of the $T(P)$ profile with the red line and the \texttt{SONORA BOBCAT} chemical equilibrium is model shown in black dashed. The second column shows the lapse-rate from the $T(P)$ profiles in blue and the local adiabatic lapse rate in black. The third column shows the chemical timescales of various gases as a function of pressure with solid lines of different colors. The pressure dependant mixing timescale is shown in the third columns with the black dashed line. The fourth column shows the iterations of the volume mixing ratios of various gases. An animated gif of this figure can be found \href{https://drive.google.com/file/d/1q9-1hKrIVHkTW__2dmXTUz0A2M3StG3G/view?usp=sharing}{here}  }
\label{fig:figiter_bd_deq}
\end{figure*}

The primary difference in the iterative scheme of the disequilibrium model is that it uses a two-step convergence method. With the user specified input parameters, the code first uses the  basic chemical equilibrium model (\S\ref{sec:chem_eq_convergence}) to converge to a chemical equilibrium atmospheric solution. This atmospheric solution is then used as the initial guess to the disequilibrium chemistry model in the second step of the iterative process. Of particular importance, is the need to recalculate the convective zones when disequilibrium chemistry is turned on.  This is important because recent work has shown that there are major differences in the location, extent and number of convective zones between the equilibrium chemistry and disequilibrium chemistry atmospheric solutions (Mukherjee et al. (2022 submitted)). The general procedure is to partially reset the convective zones before moving to disequilibrium solver. First, any upper convective zones are removed. Next, the upper boundary of the deepest convective zone is reset to slightly higher pressures (usually $\sim$5 levels in a 91-level atmosphere from 10$^{-4}$ to 200 bars). This enables the convective zones in the presence of vertical mixing to be re-calculated because we cannot assume there will be minor perturbations from the equilibrium chemistry solutions.

In order to highlight the convergence of a brown dwarf disequilibrium chemistry model, a few atmospheric iterations are shown in Figure \ref{fig:figiter_bd_deq} for a brown dwarf with {\teff} of 700 K and log(g)=5.25. The figure is similar to Figure \ref{fig:figiter_bd} except here the third column now shows the mixing (black dashed line) and chemical timescales (solid colored lines) as a function of pressure and the fourth column shows the volume mixing ratio profiles of various gases. The first row shows the atmospheric state reached after the first step of the convergence process, where a converged chemical equilibrium solution is achieved. The last row shows the final converged solution with disequilibrium chemistry. The chemical equilibrium solution from the \texttt{SONORA} grid is shown with the black dashed line in the first column. As the model iterates, chemical equilibrium $T(P)$ profile is perturbed to reach an atmospheric state where the $T(P)$ profile is visibly colder than the chemical equilibrium solution. This is a result of different atmospheric optical depths as a result of quenching of several gases in the deeper atmosphere. The chemical equilibrium solution has a single deep convective zone for this brown dwarf. However, in the iteration shown in the middle row and last row, a second convective zone also develops when disequilibrium chemistry is treated self-consistently. This shows that chemical disequilibrium also impacts the location and number of convective zones in the atmosphere. As a second convective zone develops in the middle and the third row, a decrease in mixing timescale can also be seen in that pressure range of the convective zone because convective zones are assumed to be much more efficient in mixing and thus have higher {\kzz} and lower {\tmix} than radiative zones. Comparing the fourth column in the first and the last row shows that mixing causes several orders of magnitude change in the abundance of {\co} in the upper atmosphere when compared to chemical equilibrium solutions. With these examples, we conclude our detailed discussion on the methodology of the model and move on to the model benchmarking analysis.

\section{Benchmarking Analysis}\label{sec:benchmark}

\begin{figure*}
  \centering
  \includegraphics[width=1\textwidth]{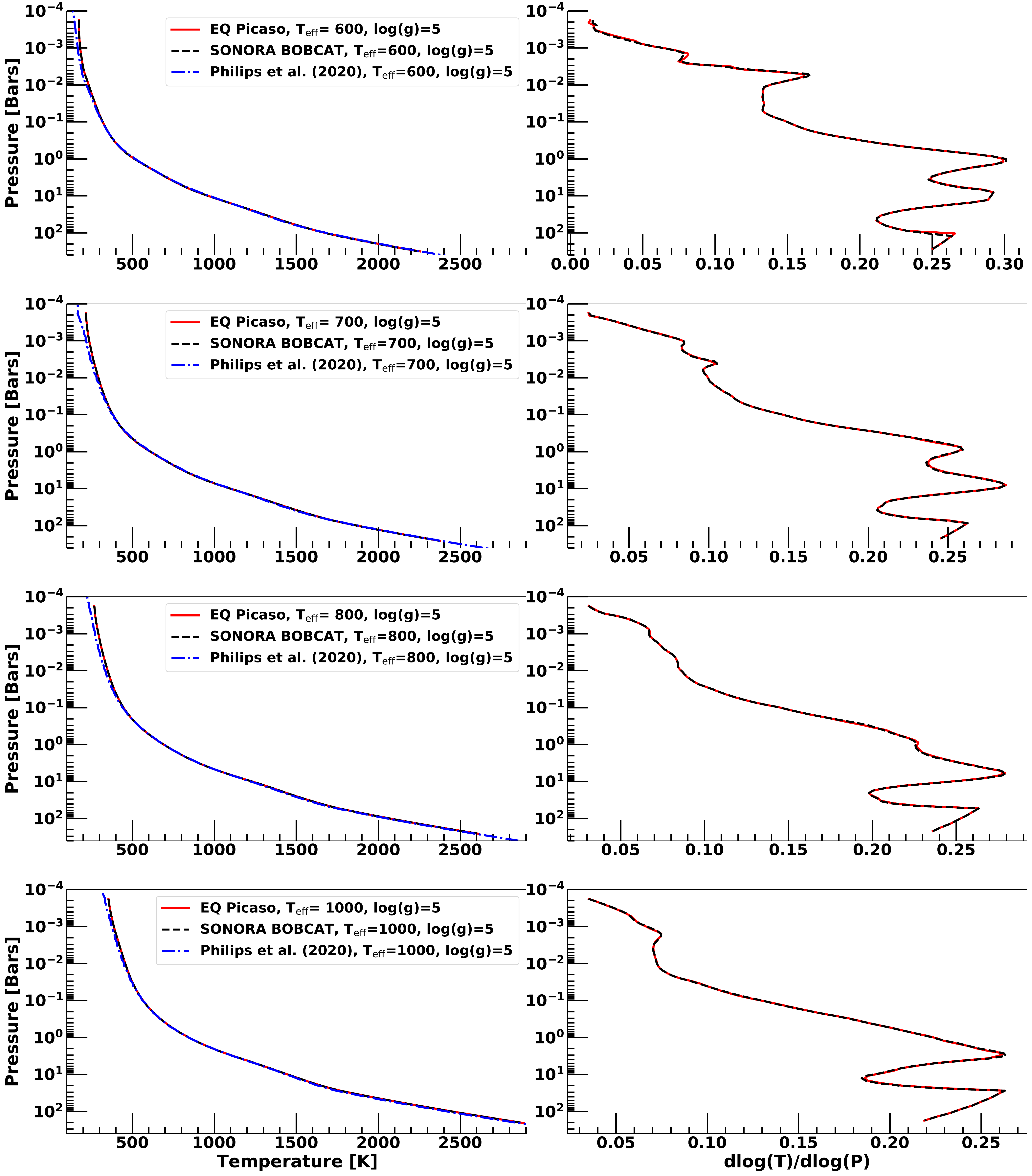}
  \caption{Benchmarking tests of the Python based-equilibrium chemistry developed in this work with the published chemical equilibrium models from \citet{marley21} and \citet{Philips20} are shown here. The benchmarking has been shown in the form of the converged $T(P)$ profiles predicted by each code for four different {\teff} values at log(g)=5 in the left panels while the pressure-derivatives of the $T(P)$ profiles also have been shown in the right side panels for each case to show that the convective and radiative zones predicted by both the codes match excellently. }
\label{fig:figbenchmark1}
\end{figure*}

\begin{figure*}
  \centering
  \includegraphics[width=1\textwidth]{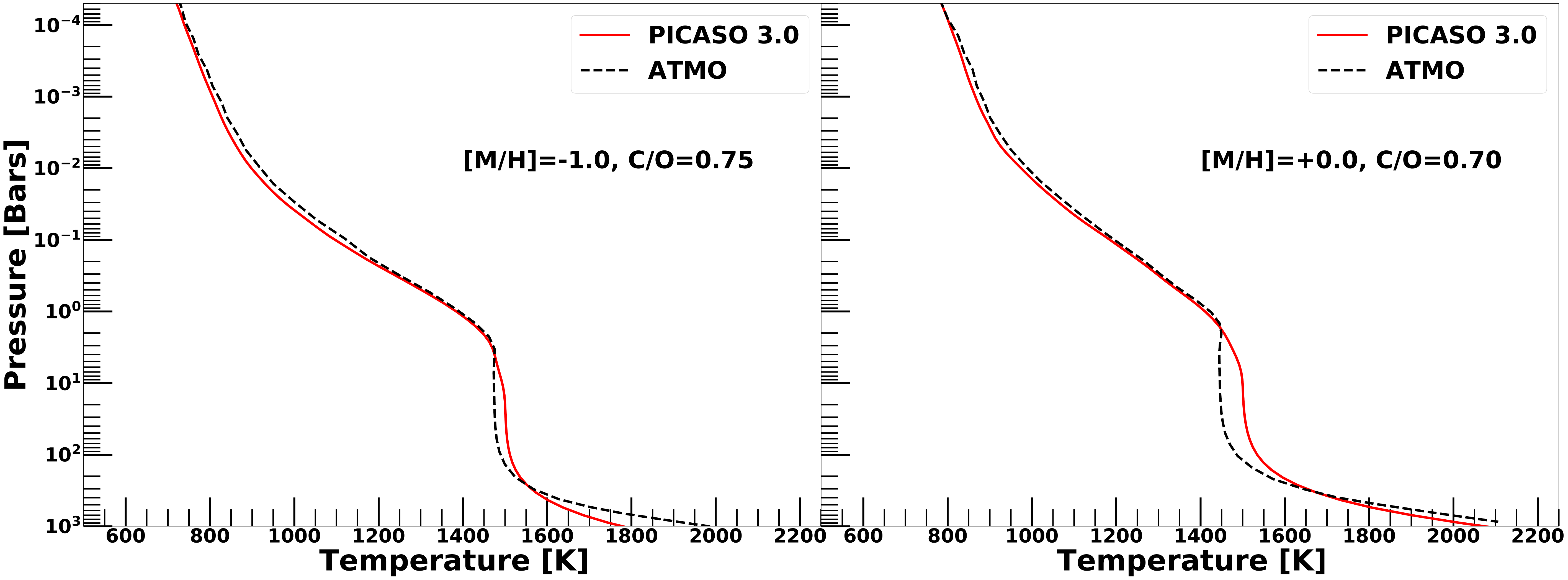}
  \caption{$T(P)$ profile computed using \texttt{PICASO 3.0} for WASP-25 b is shown with the red solid line compared with the \texttt{ATMO} models presented in \citet{goyal2020} with dashed black lines. The left panel shows the comparison for sub-solar metallicity of 0.1$\times$solar whereas the right panel shows comparisons between solar metallicity models. The same system parameters were used to compute both the models and they agree with each other with less than 3\% differences.}
\label{fig:figbenchmark4}
\end{figure*}

\begin{figure*}
  \centering
  \includegraphics[width=1\textwidth]{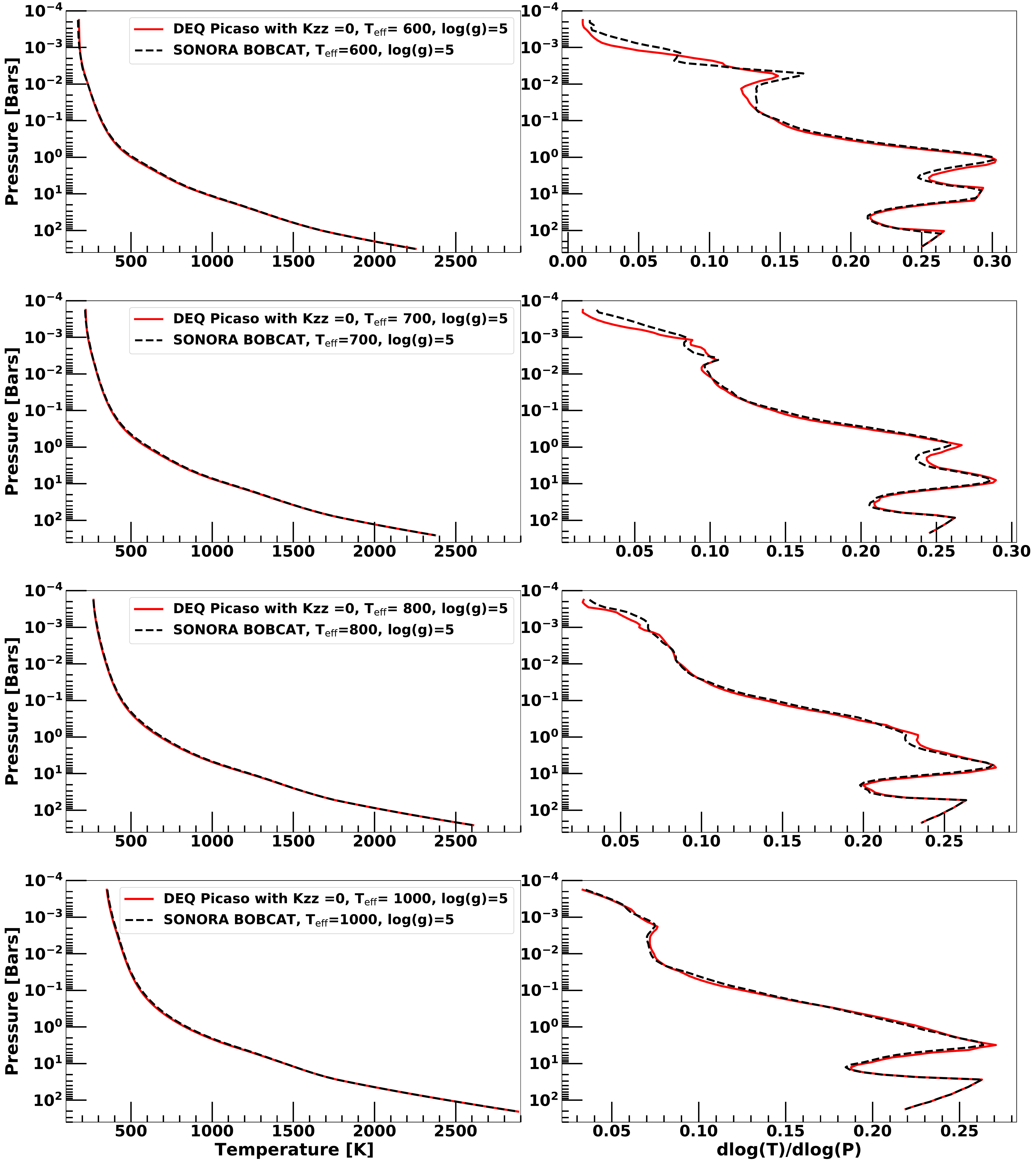}
  \caption{Benchmarking tests of the Python based-Disequilibrium code used in this work by assuming {\kzz}= 0 with the published chemical equilibrium models from \citet{marley21} are shown here. The benchmarking has been shown in the form of the converged $T(P)$ profiles predicted by each code for four different {\teff} values at log(g)=5 in the left panels while the pressure-derivatives of the $T(P)$ profiles also have been shown in the right side panels for each case to show that the convective and radiative zones predicted by both the codes match excellently. }
\label{fig:figbenchmark2}
\end{figure*}

\begin{figure*}
  \centering
  \includegraphics[width=1\textwidth]{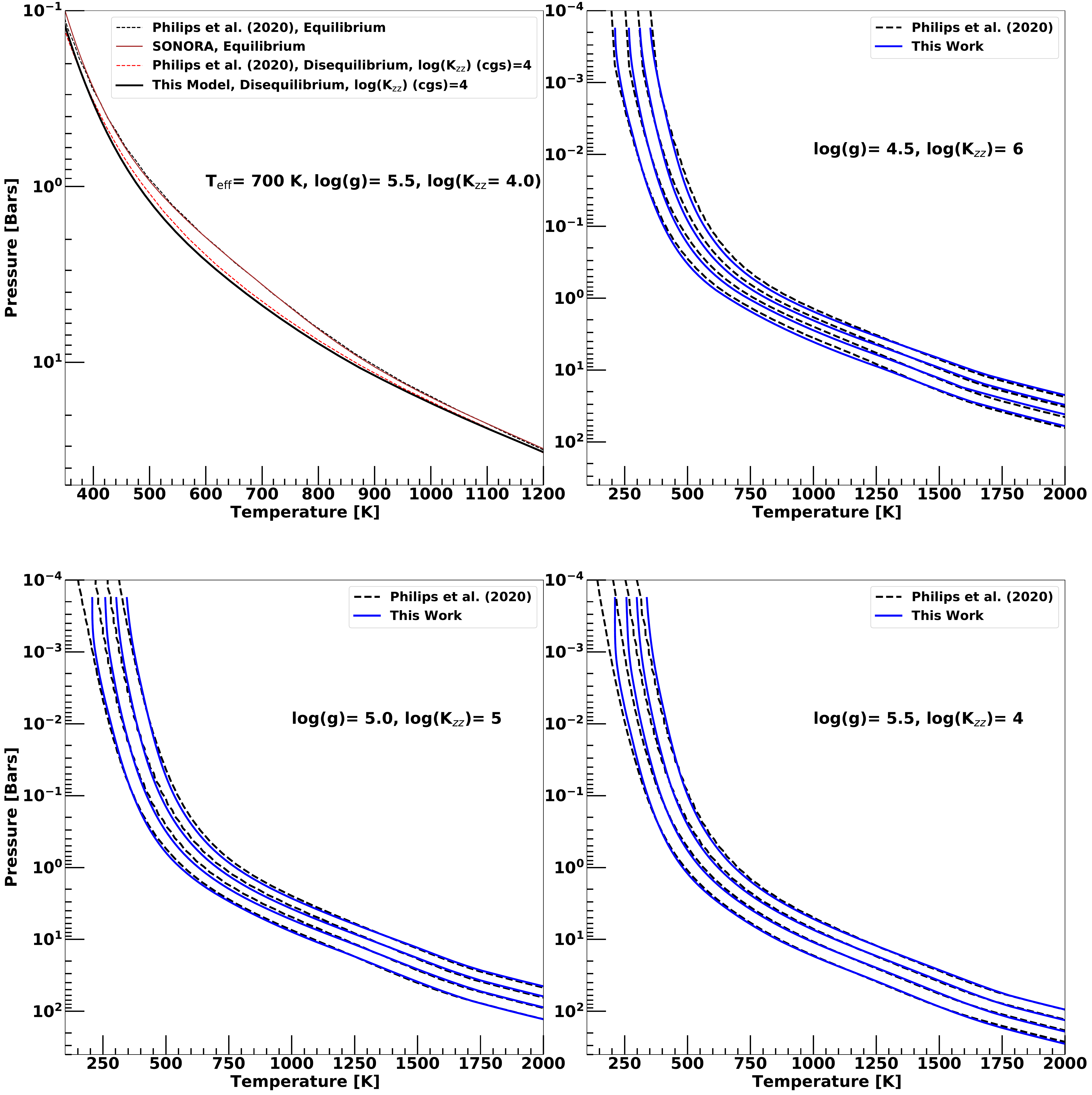}
  \caption{Benchmarking tests of the Python based-Disequilibrium code used in this work by assuming non-zero {\kzz} values with the published chemical disequilibrium models from \citet{Philips20} are shown here. The benchmarking has been shown in the form of the converged $T(P)$ profiles predicted by each code for four different {\teff} values and three different log(g) values in the three panels. The $T(P)$ profiles match well between the two independent models with different sets of opacities and totally different code methodologies.}
\label{fig:figbenchmark3}
\end{figure*}

We have benchmarked this model in four ways in order to ensure all the described methodology is functioning as expected: 1) \texttt{PICASO 3.0} chemical equilibrium non-irradiated functionality vs. \texttt{SONORA BOBCAT} grid from \citet{marley21}, 2) \texttt{PICASO 3.0} chemical equilibrium non-irradiated functionality vs. \texttt{ATMO 2020} code from \citet{Philips20}, 3) \texttt{PICASO 3.0} chemical equilibrium irradiated functionality vs. \texttt{ATMO} code from \citep{goyal2020}, and 4) \texttt{PICASO 3.0} chemical disequilibrium non-irradiated functionality vs. \texttt{ATMO 2020} code from \citet{Philips20}. 

%In order to benchmark the version of the model with chemical equilibrium, we compare \texttt{PICASO 3.0} with atmospheric models from two different grids -- the \texttt{SONORA BOBCAT} grid from \citet{marley21} and the \texttt{ATMO 2020} chemical equilibrium model grid from \citet{Philips20}. 
\subsection{Benchmarking chemical equilibrium non-irradiated atmospheres}
Figure \ref{fig:figbenchmark1} shows the comparison between the three chemical equilibrium non-irradiated models. The models described in this paper are shown in red, the \texttt{BOBCAT} models are shown with a black dashed line, and the blue dot-dashed line show the models from \citet{Philips20}. The left columns show a comparison between the $T(P)$ profiles of the three models. The right column shows a comparison between the lapse-rate of the profiles. Included in the benchmarking is four different log(g) = 5 brown dwarf models with {\teff} values of 600 K, 700 K, 800 K, and 1000 K, which covers significant chemical transitions of carbon species. \texttt{PICASO 3.0} and the models from \citep{marley21} match  well with $T(P)$ profiles  differences smaller than 0.1\%. The lapse-rates from this model also match excellently with those from \citet{marley21}. Given \texttt{PIASO}'s heritage in the EGP model used to compute the grid in \citet{marley21}, this is validation that the update to Python from Fortran did not introduce numerical issues. The disagreement between  \texttt{PICASO 3.0} and the chemical equilibrium models from the \texttt{ATMO 2020} grid are typically smaller than 1\% in the deeper atmosphere below 10 mbars. At pressures less than 10 mbars, the differences between the two models are  between 1\% to 5\%. These differences are considered minor given the independent code setup, opacity calculations, and chemistry routines. For example, models presented in \citet{Philips20} use the ``on--the--fly" opacity mixing method described in \S\ref{sec:onthefly} even when modeling atmospheres with chemical equilibrium whereas \texttt{PICASO 3.0} uses interpolations of pre-mixed opacities for iterations of chemical equilibrium models. Differences such as these along with differences in opacity line lists used by the two models could account for the 1\% to 5\% differences in $T(P)$ profiles in the upper atmosphere. 

\subsection{Benchmarking chemical equilibrium irradiated atmospheres}
We also benchmark the equilibrium chemistry version of the code by comparing models of irradiated exoplanets calculated using \texttt{PICASO 3.0} with the published \texttt{ATMO} models presented in \citet{goyal2020}. We use the RCTE models for WASP-25 b \citep{enoch11,Southworth_2014} presented in \citep{goyal2020} to perform this benchmarking.  WASP-25 b is a hot Saturn with an estimated equilibrium Temperature of $\sim$ 1210 K (assuming 0 albedo) \citep{enoch11,Southworth_2014}. We use the same system parameters for WASP-25 b as are used in the models presented in \citet{goyal2020} and detailed in the Appendix section of their article. Figure \ref{fig:figbenchmark4} shows the comparison between the two models at sub-solar and solar metallicities with a super-solar C/O ratio. The models generally agree for both the metallicities. For the sub-solar metallicity models, the agreement is better than 1\% across all the pressures. For the solar metallicity models, the maximum disagreement between the models is $\sim$ 3\%. These minor differences were also found between the \texttt{PICASO 3.0} and \texttt{ATMO 2020} models for brown dwarfs. As described above, these minor differences  can be attributed to different opacities and numerical methodologies used. 

\subsection{Benchmarking non-irradiated atmospheres with diseequilibrium chemistry}
In order to benchmark the version of the model with disequilibrium chemistry we have done two types of comparisons. First, we set {\kzz} = 0 in our disequilibrium chemistry model and compare to the results of our equilibrium chemistry model from the \texttt{SONORA} grid. Second,  we  benchmark our disequilibrium chemistry model with results from the \texttt{ATMO 2020} grid.

In the first test, if {\kzz} is assumed to be zero,  {\tmix} becomes infinitely large according to Equation \ref{eq:tmix}. Therefore, in principle, none of the gases will be quenched and their volume mixing ratios will follow equilibrium chemistry throughout the atmosphere. Therefore, a disequilibrium chemistry model run with {\kzz} = 0 must produce the same result as the equilibrium chemistry models. Of particular importance, this test checks if our ``on--the--fly" mixing routines, and double-iterative routine are performing as expected. Figure \ref{fig:figbenchmark2} shows the benchmarking between our disequilibrium chemistry model with {\kzz} = 0 and the equilibrium chemistry models from the \texttt{SONORA BOBCAT} grid. The left column shows the comparison of the $T(P)$ profiles between the two models and the right column shows the comparison of the lapse-rates between the two models. Each row from the top to bottom shows log(g) = 5.0 brown dwarf models with {\teff} of 600, 700, 800, and 1000 K. The two models match with differences $\le$ 1\% level. The small $\le$ 1\% deviations are due to the slight inaccuracies in the ``on--the--fly" mixing technique \citep{amundsen17}, which is sensitive to number of wavenumber bins,  versus the pre-weighted technique (see  \S\ref{sec:onthefly}).

In the second test, we directly benchmark our disequilibrium chemistry model with disequilibrium chemistry models from \citet{Philips20}. Disequilibrium chemistry models in \citet{Philips20} assume a constant-with-pressure {\kzz} for the atmospheres. Therefore, we use a constant {\kzz} in our benchmarking test and show the results in Figure \ref{fig:figbenchmark3}. For a brown dwarf with {\teff} of 700 K and log(g) of 5.5, the top left panel in Figure \ref{fig:figbenchmark3} shows: 1) the differences that arise in the $T(P)$ profiles when disequilibrium is turned on for both the models from \citet{Philips20} and \texttt{PICASO 3.0}, and 2) the differences that arise between  \citet{Philips20} and \texttt{PICASO 3.0}. Both the equilibrium chemistry and disequilibrium chemistry models show excellent matches with among themselves which is an excellent benchmarking demonstration of our methods and codes. Additionally, this also shows the impact of disequilibrium chemistry on the $T(P)$ profiles of this representative brown dwarfs, which results in $\sim$50~K colder $T(P)$ profiles compared to equilibrium chemistry models. 

The rest of the panels in Figure \ref{fig:figbenchmark3} show comparisons between various disequilibrium chemistry $T(P)$ profiles with different {\kzz} and gravity values produced with our model with the grid of models from \citet{Philips20}. Our disequilibrium models generally agree well within 5\% levels in the deeper atmosphere with models from \citet{Philips20}. However, the disagreements are higher in the much lower pressure upper atmosphere which was also the case for the equilibrium chemistry model comparisons. The small disagreements between the models are mainly due to two reasons -- 1) \citet{Philips20} use chemical kinetics models for calculation of quenched abundances whereas we use the ``quench-time" approximation (\S\ref{sec:deq_chem}) to do so and 2) differences between the accuracy of the ``on--the--fly" mixing method used between the two models due to differences in number of wavelength bins used. These disagreements show the uncertainty in state--of--the--art models which perhaps can be explained in future with comparisons with high signal-to-noise data.

\section{Modeling Recommendations}\label{sec:recommendations}
The code is generally well-behaved across the parameter space of interest for brown dwarfs and exoplanets. However, iterative schemes are sometimes notoriously tricky to converge. Therefore, here we outline recommendations regarding the use of code such that users can get meaningful and accurate results. These recommendations are implemented in the publicly available code tutorials available via the \texttt{PICASO} documentation page \footnote{https://natashabatalha.github.io/picaso/tutorials.html released with publication acceptance}.

\subsection{Choosing Model Pressure Grid}

We recommend using typically 50-90 atmospheric pressure layers, corresponding to 51-91 pressure levels. In general, higher number of pressure layers  increases the computational time required for convergence substantially and lower number of layers makes the atmospheric grid too coarse for an accurate calculation. The maximum and minimum values of the pressure grid are also ultimately chosen by the user. While the choice of the minimum pressure can be made somewhat arbitrarily, we recommend that the model not be run at pressures  lower than 10$^{-6}$. This is the  lowest pressure for which chemistry and opacities are computed (shown in Figure \ref{fig:figabun} and \ref{fig:figpremixopa}). Another uncertainty from the double-Gauss correlated-k approach can arise from running the model at very low pressures ($\le$ 10$^{-4}$ bars). As the molecular lines become very narrow at such low pressures, most of the opacities resides at very high values of $G(\kappa)$ (e.g. between 0.995-1.0) in the k-distribution. Therefore, these opacities are not taken into account even with the double-Gauss method. This can make the radiative transfer at such low pressures inaccurate.

The maximum pressure of the atmospheric model needs to chosen carefully in the case of giant planet atmospheres and brown dwarfs. If the maximum pressure of the atmospheric model is too low, then the model can become transparent to the deepest atmospheric layer, especially in wavelengths with little gaseous opacities (e.g. optical wavelengths $\lessapprox1\mu$m). This would result in an  inaccurate $T(P)$ profile. A good way to check if the profile was run with low maximum pressure is to plot the wavelength dependent brightness temperature of the converged model along with the converged temperature of the deepest atmospheric layer. If the brightness temperature at any wavelength exceeds the brightness temperature associated with the bottom temperature, then the pressure grid is not well-suited for the calculation. However, if the brightness temperature is smaller than this bottom temperature at all wavelength, then the pressure grid is well-suited for the calculation. In the first scenario, the pressure grid needs to be extended to higher pressures. The problem with this approach to choose a pressure-grid is that it can only be done after running the model once. A more practical alternative to this approach is to find a comparable model from the \texttt{SONORA BOBCAT} \footnote{\href{https://zenodo.org/record/5063476\#.YkX8tjfMJhE}{Link to published SONORA BOBCAT models}} to the temperature, gravity parameter space wanted and use the maximum pressure corresponding to that model. Also, the highest pressure for which chemistry and opacities are computed (shown in Figure \ref{fig:figabun} and \ref{fig:figpremixopa}) is 3000 bars. Therefore, runs with pressure grids extending to higher pressures than this will be inaccurate as they would require linear extrapolation of both gas abundances and opacities.

When computing  disequilibrium chemistry runs, the user must take another aspect into consideration while choosing the maximum pressure of the atmospheric pressure grid: the quench levels of various gases. If the maximum pressure of the atmospheric grid is too low such that gases are expected to quench at deeper parts of the atmosphere than the pressure grid, then the model run will be incorrect. Therefore, users must check the quench levels of various gases after performing a disequilibrium chemistry calculation with the model. If the quench levels of any of the gases are at the last pressure level, then the user needs to increase the maximum pressure of the run to get a correct converged solution.

\subsection{Choosing the Initial Guess T(P) Profile}
Climate models that leverage an iterative scheme require a first guess of the $T(P)$ profile. This guess $T(P)$ profile is then iterated to reach the converged solution. Even though we  used isothermal $T(P)$ profiles as our first guess in both Figure \ref{fig:figiter_bd} and \ref{fig:figiter_exo}, it is usually preferable to start with an initial guess that is close to the expected solution.  Using  simple profiles like isothermal profiles can lead to significant increase in run times and in the worst cases  also lead to solutions which do not converge. A far better alternative to isothermal profiles, are publicly available model grids like \texttt{SONORA BOBCAT} or the \citet{Philips20} models. For exoplanets, we recommend using parametrized $T(P)$ profiles  from, for example \citet{guillot10}, as a first guess. Functionalities to browse these profiles directly are already available in \texttt{PICASO}, via the \href{https://natashabatalha.github.io/picaso/picaso.html?highlight=guillot_pt#picaso.justdoit.inputs.guillot_pt}{\texttt{guillot\_pt()}}, and \href{https://natashabatalha.github.io/picaso/picaso.html?highlight=sonora#picaso.justdoit.inputs.sonora}{\texttt{sonora()}}
functions \footnote{also see various tutorials e.g. \href{https://natashabatalha.github.io/picaso/notebooks/6_BrownDwarfs.html\#Download-and-Query-from-Sonora-Profile-Grid}{Download and Query from Sonora Profile Grid}, and \href{https://natashabatalha.github.io/picaso/notebooks/FAQs.html\#How-do-I-access-the-pressure-temperature-profile-parameterizations?}{Accessing the pressure-temperature profile parameterizations}}.  

Another first guess required to run the model is the extent of the deepest convective zone. Specifically, the user needs to specify a guess of the pressure level of the radiative-convective boundary in the atmosphere which the code will then modify using methods described in \S\ref{sec:convec}. Even though this guess is modified within the iteration, we recommend always setting the bottom 3-5 levels to be convective first. If the user sets too many layers to be convective, the converged solutions may be inaccurate because the model is only equipped to grow and merge convective zones, not shrink them. Therefore, the best practice is to start with 3-5 bottom convective levels. If the code runs into convergence issues after doing so, we recommended increasing this number slowly to check if it helps converge the solution.

\subsection{Convergence}\label{sec:convergence_rec}

In most cloud-free cases, the code will converge. However, there are a few  known cases where extra steps are needed to reach a converged solution. For equilibrium chemistry brown dwarf models with {\teff} between 1500 K--1700 K, an oscillating behaviour is  seen in the iterations of the $T(P)$ profile. In these cases, the $T(P)$ profile oscillates between two profiles with a constant temperature offset. This behaviour is not unexpected and is caused by the sharp ``cliff" in gaseous opacity at these effective temperatures. This can be seen in the cross-section maps in Figure \ref{fig:figpremixopa}. All the four panels in Figure \ref{fig:figpremixopa} show a  sharp change in gaseous opacity around the 900 K--1700 K temperature range. The overlap of this cliff with the converged brown dwarf $T(P)$ profile at {\teff}=1700 K can also be seen in Figure \ref{fig:figpremixopa}. As already mentioned in \S\ref{sec:opa}, this sharp opacity ``cliff" makes the convergence difficult around this temperature range. In order to overcome this, we recommend performing multiple runs for the same object by recycling the unconverged final $T(P)$ profile of each run as an initial guess profile for the next run until a converged solution is reached. 

A similar convergence issue also occurs for objects with {\teff} $<$ 300 K. Figure \ref{fig:figabun} bottom left panel shows how {\water} abundance in the vapor phase falls drastically due to {\water} condensation at such low temperatures. This sharp drop in the {\water} abundance causes the $T(P)$ profiles of these cold objects at the low pressure regions to oscillate between multiple possible solutions. This behaviour can be seen in the upper parts of the {\teff}=300 K $T(P)$ profile in Figure \ref{fig:figpremixopa}. We recommend two solutions to this problem: 1) either set the minimum pressure of the model pressure grid  to no less than 1 mbar,  or 2) use a \texttt{SONORA} profile as an initial guess and recycle the unconverged profile multiple times ($\sim$ 4 times) to reach a converged state. The first recommendation helps the user  exclude the part of the $T(P)$ profile which causes the instability in the convergence. As these objects are cold, the excluded pressure region is extremely cold ($<$ 120 K), which means that there is practically no contribution to the observables (e.g. total emergent flux).

\section{Calculating Observables}

After using the radiative-convective equilibrium model to calculate the atmospheric structure of a brown dwarf or an exoplanet, the user may want to compute various observables. \texttt{PICASO} has been updated to include the capability of calculating the 1D and 3D thermal emission spectrum, the 1D transmission spectrum, and the 1D and 3D reflection spectroscopy, and phase curves. With a 1D $T(P)$ profile, users will likely want to produce 1D higher resolution thermal emission spectra, reflection spectra, and/or transmission spectra of the planet or brown dwarf. These calculated observables can be directly compared with observational data.

\begin{figure*}
  \centering
  \includegraphics[width=1\textwidth]{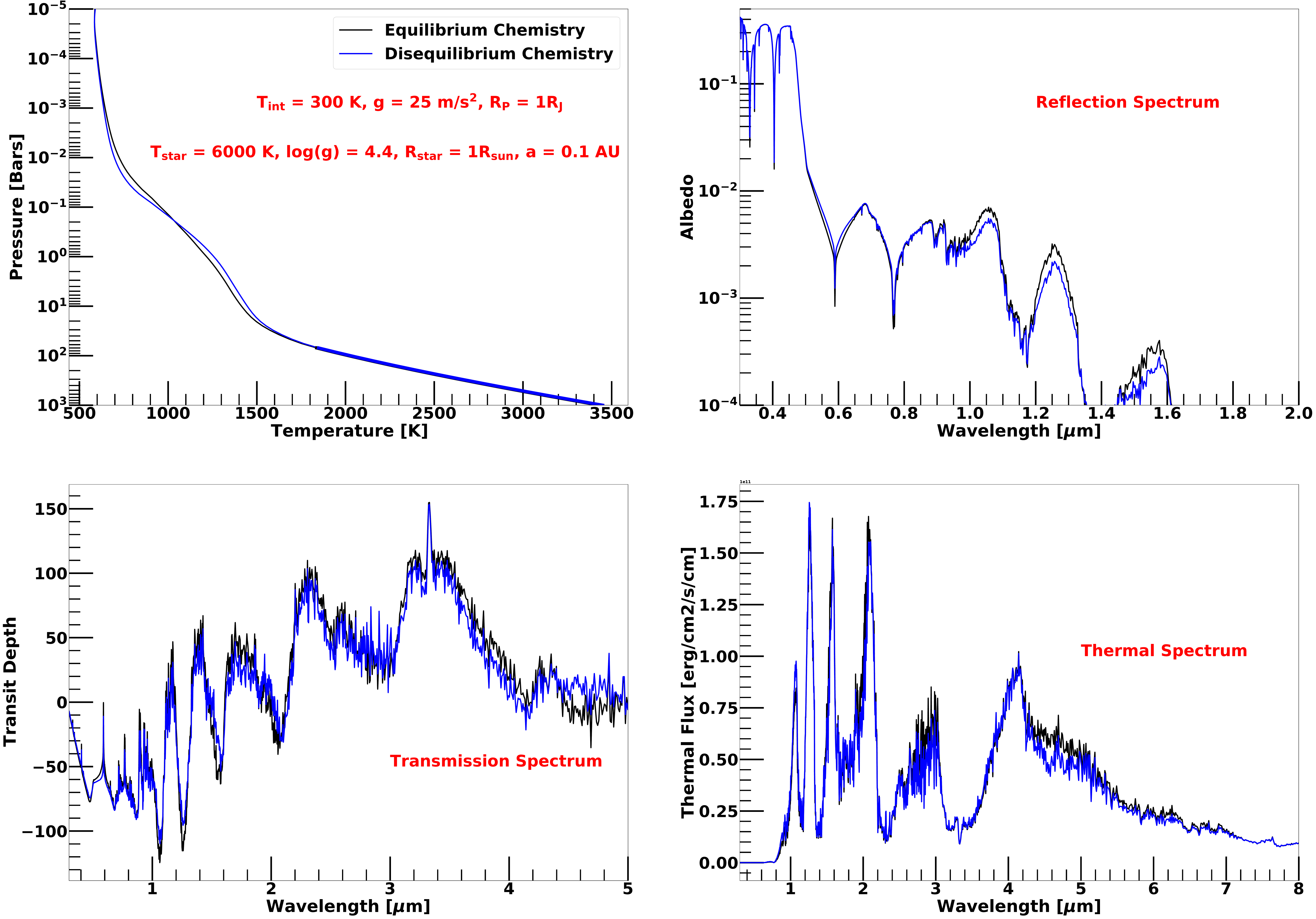}
  \caption{Using our atmospheric model to calculate observables. {\bf Top left} panel shows the converged $T(P)$ of an exoplanet calculated using our model with (blue) and without (black) disequilibrium chemistry. {\bf Top right}, {\bf bottom left} and {\bf bottom right} panel shows the albedo spectrum, transmission spectrum, and thermal spectrum at a spectral resolution of R=300 of the exoplanet calculated using radiative transfer tools available in \texttt{PICASO}.}
\label{fig:figobservable}
\end{figure*}

Figure \ref{fig:figobservable} shows an example of this with an exoplanet. The exoplanet has been assumed to be a Jupiter mass and size planet with {\tint}=300 K at a distance of 0.1 AU around a Sun like star. The radiative-convective equilibrium model described in this work has been used to calculate the planet-wide average atmospheric structure of this hypothetical planet once with chemical equilibrium and then again with chemical disequilibrium. The top left panel in Figure \ref{fig:figobservable} shows the $T(P)$ profile of this planet in each case. It is clear that self-consistent treatment of disequilibrium chemistry also impact the $T(P)$ profile of irradiated planets compared to equilibrium chemistry models. We have then used the 1D radiative transfer routines of \texttt{PICASO} to compute the reflection albedo spectrum of this exoplanet. This albedo spectrum has been shown in Figure \ref{fig:figobservable} top right panel. The planet is not very reflective after 0.5 $\mu$m, after which brightness from Rayleigh scattering tapers off. The transmission spectra of the planet is also modeled with \texttt{PICASO} and is shown in Figure \ref{fig:figobservable} bottom left panel. This panel shows that the presence of disequilibrium chemistry can be clearly detected for a hot Jupiter like this with its transmission spectra between 3-5 $\mu$m. The thermal spectra are shown in the bottom right panel and the presence of disequilibrium chemistry in hot Jupiters also can be easily detected with their thermal spectra in the M-band, or in various NIRSpec and/or NIRCam JWST modes. Note that for this illustrative example, we have used the planet-wide average computed $T(P)$ profile for all these viewing geometries. However, a more self-consistent method to compute these viewing geometries would be to compute a day-side averaged profile for use in the zero-phase reflection and thermal spectrum, and a planet-wide average for the transmission spectrum. 

\texttt{PICASO} has also been equipped with simple modules to couple the atmospheric models produced from this code with the evolutionary models from \citet{marley21} to calculate absolute Vega magnitudes of the modeled object in any filter of choice. The atmospheric model presented in this paper used with these radiative transfer tool are already available in \texttt{PICASO} and can be immensely helpful in modeling or planning observations using models. 

\section{Future Improvements}\label{sec:improve}

Currently, our publicly released model has the capability to model non-cloudy brown dwarfs and exoplanets with equilibrium chemistry and disequilibrium chemistry. However, several future improvements of the model are needed to enhance its capability to capture a larger parameter space. Here we briefly discuss these needed future improvements. We note that all \texttt{PICASO} development is public, and open to community involvement. 

\subsection{Clouds}
Our iterative climate model currently does not include the capability to treat atmospheric clouds self-consistently. The \texttt{EGP} code already includes condensation clouds with the \citet{ackerman2001cloud} model and has been used to model cloudy L-dwarf atmospheres \citep[e.g.][]{cushing08,stephens09}. Clouds can become  important opacity sources in brown dwarf and exoplanetary atmospheres due to their strong scattering tendencies. This also has large effects on the $T(P)$ profile of these atmospheres \citep{morley14water,morley2012neglected}. We plan to couple this model with the Python-based cloud model \texttt{VIRGA} \citep{virga}. Recently, \texttt{VIRGA} has been updated to include variable sedimentation efficiency (f$_{\rm sed}$) with height \citep{rooney21}. We will couple this updated \texttt{VIRGA} model with our model so that our models can be used to model hotter L-dwarfs which are generally assumed to be cloudy and also much colder Y-dwarfs with {\water} clouds. This improvement will also enable us to apply our model to cloudy exoplanets and brown dwarfs. 

\subsection{1D Chemical Kinetics Model}

We use the quench time approximation to model the effects of vertical mixing on the chemical abundance profiles in this model. This approximation is based on the parametrized mixing timescales from \citet{Zahnle14} outlined in \S\ref{sec:deq_chem}. However, more robust 1D chemical kinetics models of treating disequilibrium chemistry caused by vertical mixing are now open-sourced like the \texttt{VULCAN} model \citep{tsai17,tsai21}. We plan to  couple our Python model with the \texttt{VULCAN} chemical kinetics model to increase the flexibility of our model. %This coupling will allow us to not rely on the quench time approximation for disequilibrium chemistry calculations. 
This will make our disequilibrium chemistry models more robust. Moreover, \texttt{VULCAN} also includes the capability to treat stellar irradiation induced photochemistry. This will lead to a significant improvement for applicability of our model to  irradiated exoplanets where photochemistry can largely impact atmospheric chemistry especially at lower pressures. This update will allow us to explore the impact of photochemistry on the atmospheric structure of exoplanets. We have already made progress for this coupling but still need to test and benchmark several aspects of the coupled code before this update is publicly available for use.

\subsection{Time Evolution Version}
Recently, \citet{mayorga21,robinson14} enhanced the \texttt{EGP} model to \texttt{EGP+} by including a time-stepping version of the code which can model the dynamic temporal response of the atmosphere to time-varying physical conditions like stellar irradiation changing with time. This is especially relevant to planets with highly eccentric orbits as the stellar irradiation can change by a large amount within one orbital time-period for these planets. We plan to adapt this improvement within our code in the near future as well. Studying the atmospheres of  eccentric exoplanets is a  promising research area and a subject of future JWST observing campaigns.

\section{Conclusions and Summary}\label{sec:summary}

We  presented a new open-source Python based 1D radiative--convective equilibrium model as part of the \texttt{PICASO} package. This code derives its heritage from the \texttt{EGP} code developed by \citet{marley1999thermal} based upon a Titan atmosphere model developed by \citet{mckay1989thermal}. The \texttt{EGP} code has been used to model brown dwarf and exoplanetary atmospheres for almost three decades now. The model is applicable to H-dominated atmospheres of both brown dwarfs and irradiated exoplanets. The model includes the capability to do calculations with both equilibrium chemistry and disequilibrium chemistry due to vertical mixing. The model includes options to use {\kzz} values constant with height while performing disequilibrium chemistry runs. We have also included the capability to use self-consistent prescriptions of {\kzz} within the model where the {\kzz} will also iterate along with the $T(P)$ profile and atmospheric chemistry to ultimately reach a converged atmospheric state.

We have benchmarked this model with publicly available models from the \texttt{SONORA BOBCAT} grid and also with results from an independent model used by \citet{Philips20} to produce the \texttt{ATMO 2020} atmospheric grid. For irradiated planets, the \texttt{PICASO 3.0} models were benchmarked against the \texttt{ATMO} grid presented in \citet{goyal2020}, using the hot Saturn WASP 25-b as a test case. The chemical equilibrium version of the model was benchmarked both with models from the \texttt{SONORA} grid, \texttt{ATMO 2020}, and the \texttt{ATMO} models for WASP 25-b. The chemical disequilibrium models were benchmarked against models from the disequilibrium chemistry atmospheric models from the \texttt{ATMO 2020} grid of models. This benchmarking analysis showed excellent agreement with \texttt{PICASO}.

Our model is open-source and publicly available for the community (we include several in text links to code throughout this manuscript). Additionally, we outlined many recommendations in this work for proper usage of the model. This includes recommendations on choosing the atmospheric pressure grid for a science object, choosing the initial guess $T(P)$ profile for a particular run and also the various parts of the parameter space where the model is known to face convergence issues. We also included ways of resolving these convergence issues.

We will also release tutorials to apply this model for various science cases with the code, upon paper acceptance. We plan to actively develop this model further to include clouds, couple it with 1D chemical kinetics codes for better robustness and also include time dependent effects like variable stellar irradiation within our model.

\section{Acknowledgments}
SM thanks the UC Regents Fellowship award for supporting him for this work. NEB acknowledges support from NASA Astrophysics Division.  NEB and JJF acknowledge support from NASA’S Interdisciplinary Consortia for Astrobiology Research (NNH19ZDA001N-ICAR) under award number 19-ICAR19\_2-0041. JJF and MSM acknowledge the support of NASA XRP grant 80NSSC19K0446. We thank the anonymous referee for helping us to improve the manuscript significantly.
 
{\it Software:} PICASO \citep{batalha19}\footnote{upon acceptance we will formally release PICASO 3.0 and update this with the Zenodo link}, pandas \citep{mckinney2010data}, NumPy \citep{walt2011numpy}, IPython \citep{perez2007ipython}, Jupyter \citep{kluyver2016jupyter}, matplotlib \citep{Hunter:2007}

\bibliography{sample_arxiv}{}
\bibliographystyle{apj}



\end{document}